%% file: main_cvpr_cr_4arxiv_with_supp.tex
\documentclass[10pt,twocolumn,letterpaper]{article}

\usepackage{cvpr}  

\definecolor{cvprblue}{rgb}{0.21,0.49,0.74}
\usepackage[pagebackref,breaklinks,colorlinks,allcolors=cvprblue]{hyperref}
\usepackage{bm}
\usepackage{xcolor}
\usepackage{times}
\usepackage{graphicx}
\usepackage{bmpsize}
\usepackage{my_macros_cvpr} 
\usepackage{amsmath}
\usepackage{nameref}
\usepackage{multirow}

\def\paperID{1271} 
\def\confName{CVPR}
\def\confYear{2025}

\vspace{-0.2cm}
\title{ Understanding multi-layered transmission matrices\vspace{-0.2cm}}

\author{Anat Levin and Marina Alterman\\
	Department of Electrical and Computer Engineering, Technion\\
	Haifa, Israel\\
}

\begin{document}
	\maketitle
%
%
%


\input{cvpr_sec_files_cr/abstract}



\input{cvpr_sec_files_cr/intro}


\input{cvpr_sec_files_cr/problem_formulation}
\input{cvpr_sec_files_cr/fittingevals}

\input{cvpr_sec_files_cr/empirical_fits}

\input{cvpr_sec_files_cr/missingcone}

\input{cvpr_sec_files_cr/boundedsupport}

\vspace{-0.2cm}\section{Discussion}\vspace{-0.2cm}
\input{cvpr_sec_files_cr/discussion}

{
	 \small
\bibliographystyle{ieeenat_fullname}
\bibliography{biblio_anat}
}

\section*{Supplementary}
\appendix
\input{cvpr_sec_files_cr/missingcone_mc}

\input{cvpr_sec_files_cr/empirical_fits_supp}

\input{cvpr_sec_files_cr/volumesampling}

\end{document}

%% file: cvpr_sec_files_cr/abstract.tex
\vspace{-0.2cm}\begin{abstract} 
	Wavefront shaping systems attempt to correct aberrations caused by tissue scattering, by placing a spatial light modulator (SLM) in the optical path and using it to reshape the light wave emitted from a target of interest deep inside the tissue. However, the field-of-view we can correct with one SLM pattern is extremely small, because inherently the aberration is caused by the 3D tissue structure while the SLM is planar. To overcome this, multi-conjugate correction systems built with multiple layered  SLMs  have been introduced, which attempt to approximate the 3D tissue structure with multiple planar aberrations.

	However, multi-conjugate systems with a large number of SLMs are not easy to construct. Therefore, it is important to understand how many SLM layers are actually needed for a good correction, and whether there are practical benefits from correction systems with a relatively small number of layers.
	 To help in the design of future systems, this paper analyzes multi-layer corrections. We show that the well-known missing cone problem which fundamentally limits the axial resolution of microscopes, turns into an advantage when designing 3D correction systems. Due to this property less information should be captured by the correction system, facilitating sparser approximations.     We also show that even if the number of available layers is insufficient for a reasonable approximation of the full 3D tissue volume, it can significantly expand the field-of-view when compared to single-layer correction  systems,  and therefore can largely accelerate their operation.

\end{abstract}\vspace{-0.2cm}

%% file: cvpr_sec_files_cr/intro.tex
\vspace{-0.3cm}\section{Introduction}\vspace{-0.2cm}
Scattering is a central challenge in multiple imaging tasks ranging from astronomy to microscopy, and biomedical imaging. It degrades our ability to resolve stars through telescopes, image into the bottom of the sea, and drive through foggy weather. Most importantly in the context of this paper, scattering severely limits our ability to image through biological tissue. This happens due to small variations in the refractive index of the tissue volume, which scatters the propagating light. 

Numerous computational optics approaches have been developed for imaging through scattering tissue. These include computational restoration algorithms which attempt to fit the observed data with various models for the aberration and the target of interest~\cite{Metzler23NeuWS,YeminyKatz2021,haim2023imageguidedcomputationalholographicwavefront,Balondrade_2024,Kang2017,Najar2024, Zhu22, Baek_2023,Jeong2018,Gil2024,Yonghyeon2022}. Also, optical systems such as a confocal microscope~\cite{ConfocalMicroscopyOverView2020} or an OCT~\cite{OCTOverview2016}, aim to filter out the scattered light and maintain only the ballistic unscattered part. However, as we attempt to see deeper into a scattering layer, ballistic light weakens significantly, and the complexity of aberration models increases. Moreover, biological target of interest such as fluorescent tissue components are very weak and as a result the scattered signal is also of a very low signal to noise ratio (SNR), further limiting digital correction models. A particularly attractive approach for scattering aberration removal is wavefront shaping, where one places a spatial light modulator (SLM) in the optical path of the microscope and uses it to correct the wavefront emitted from a point deep inside the tissue, so that light photons emerging from a single target point are brought into a single sensor point, despite the tissue aberration. Wavefront shaping is a particularly attractive solution for fluorescent imaging because the correction is done optically rather than digitally. Unlike ballistic filtering approaches, all light photons are used. 
Since the correction is done in the optical path,  the limited number of available photons is brought into a single  detector pixel and can be measured with higher SNR. As a result, wavefront shaping has the potential to revolutionize tissue imaging and allow us
to overcome the fundamental limitations of scattering, and image very deep inside biological targets at a high SNR.
 
The main limitation for wavefront shaping approaches is that the aberrations of neighboring points
inside volumetric tissue vary rapidly, and in practice a single correction mask can only allow imaging of
a small local field-of-view. This is because the scattering process happens throughout the 3D volume, while
the aberration correction only happens on a planar SLM.
Earlier demonstrations of wavefront shaping have
considered targets placed at a distance behind a thin scattering layer, and this far-field arrangement has
allowed them to correct larger regions with a single planar modulation~\cite{YeminyKatz2021,Stern:19,Daniel:19,Metzler23NeuWS}. Similarly in the
context of adaptive-optics, researchers have attempted to correct aberrations due to imperfect optics or a refractive index mismatch~\cite{Booth2014,Ji2017review,HampsonBooth21review}. These aberrations are introduced along the optical path, before the tissue, so the correction can apply to a larger area. However, in most biomedical applications the fluorescent target of interest is
embedded within the scattering volume. If our goal is to correct scattering within the tissue itself, the field-of-view which can be corrected by a single planar modulation is extremely small, usually only a few microns~\cite{SeeThroughSubmission}.

To correct larger fields of view, {\em multi conjugate} correction systems have been proposed. Such systems use multiple planar SLMs placed such that each of them is conjugate to a different plane inside the tissue volume. Each such SLM should realize a correction for the aberration happening inside a thin layer of the volume and the combination of such multi-plane corrections should approximate the 3D variation in refractive index, and allow us to correct it. \figref{fig:setup-wavefront-shaping} visualizes the concept of a multi-conjugate correction~\cite{Thaung:09,Wu:15,Laslandes:17,Furieri23}.  
In practice, multi-plane correction systems are more complicated than the simplified visualization of \figref{fig:setup-wavefront-shaping}, as one needs to add multiple relay systems between the SLMs. This is first because there are physical limitation on the minimal distance between 2 SLMs, and second because we need to revert the order of the correction layers in the inversion path~\cite{1998aoat.bookH}. This makes multi-conjugate correction systems complex and cumbersome to build. Previous experimental demonstrations have  realized corrections with no more than two SLMs. While two-layer correction systems have demonstrated improved performance compared to single layer correction, two planes clearly cannot model a thick 3D tissue structure.

Our goal in this research is to analytically study multi-slice models and attempt to answer the following questions: How many planes are required to approximate the 3D aberration of a scattering tissue volume? How does the quality of the correction degrade if we use a sparser approximation? If a construction of a practical correction system is limited to only to M layers, what is the maximum field of view that can be effectively corrected?

To this end, we use the mathematical model of a transmission matrix, which stores the propagation of coherent wavefronts from one end of a scattering volume to the other. We study how well we can approximate such transmission matrices with a multi-slice model, and how does the error depend on the  number of fitted layers.

 The straightforward answer to this question is to say that the spacing between layers should match the axial resolution of the microscope. For example, using an $\NA=0.5$ objective at wavelength $500nm$, we get that the spacing between layers should be lower than  $2\mu m$. Handling a modest tissue  thickness of, e.g., $200\mu m$ already requires $100$ slices for a valid approximation. Clearly building a multi-conjugate wavefront shaping systems with so many layers is impractical.
 
 \input{cvpr_sec_files_cr/fig_wavefrontshaping_setup} 
In this research we take inspiration from~\cite{Hua:23}, and offer a thorough analysis of the frequency content of transmission matrices, combining  the forward scattering nature of biological tissue with the structure of the missing cone.
We show that while the missing cone structure is known as one of the fundamental limitations of 3D microscopy, in the context of transmission matrices this limitation turns into an advantage.  Due to the missing cone structure, many of the frequencies of the volumetric aberration do not participate in the transmission matrix, and therefore the number of layers required for approximating a transmission matrix is way smaller than the Nyquist limit. 
However, there are no free lunches and the number of layers scales with the tissue thickness. To explain a few hundred micron thick tissue one needs a few dozen   layers, which is still too much to be realized by a physical correction system. 
\input{cvpr_sec_files_cr/fig_setup}

Given this conclusion we aim to understand if there are any practical benefits in realizing a multi-conjugate correction system with a small number of say $2-4$ layers. We show that while a small number of layers cannot well approximate the 3D structure of the tissue, it can allow us to realize a correction over a limited field-of-view. The area which can be corrected with $M$ aberration layers is significantly larger than the area which can be corrected with one layer, and therefore it can significantly accelerate sequential correction of the field of view.
We validate our findings  using real transmission matrices captured in the lab as well as synthetic transmission matrices simulated using accurate wave propagation  models.

%

%% file: cvpr_sec_files_cr/fig_wavefrontshaping_setup.tex

\vspace{-0.1cm}\begin{figure}[t!]
	\begin{center}
\begin{tabular}{ll}
	\includegraphics[width=0.12\linewidth]{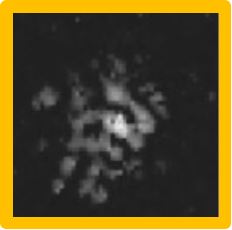} & \multirow{2}{*}[1cm]{\hspace{2.4 cm} \includegraphics[width=0.52\linewidth]{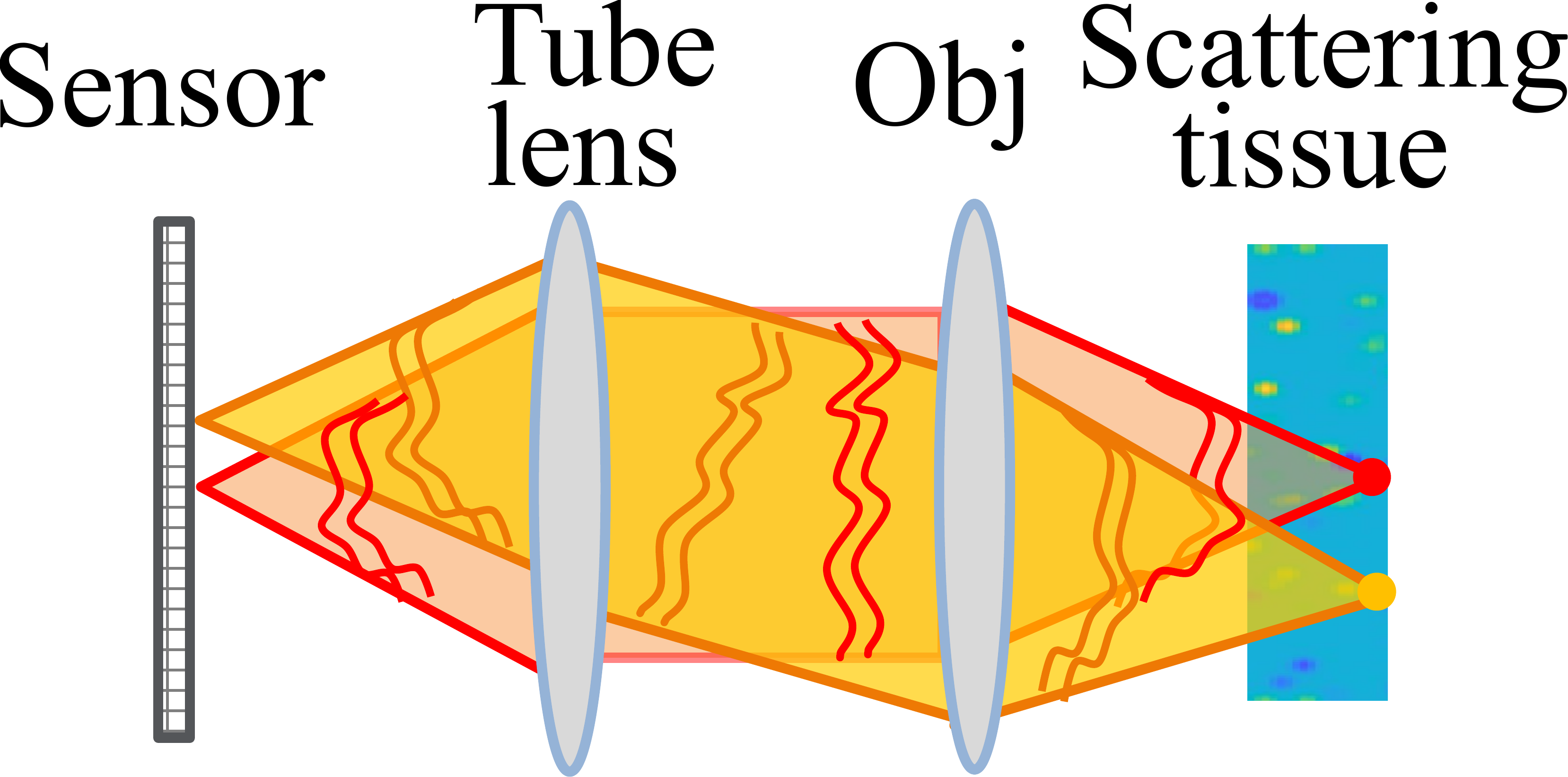}} \\[-0.7ex]
	\includegraphics[width=0.12\linewidth]{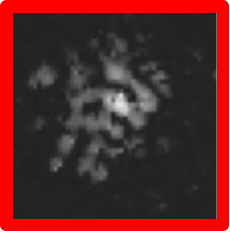} &                   \\
	\includegraphics[height=0.12\linewidth]{figs/setup/cam_nofocus_orange.pdf} & \multirow{2}{*}[0.5cm]{\includegraphics[width=0.8\linewidth]{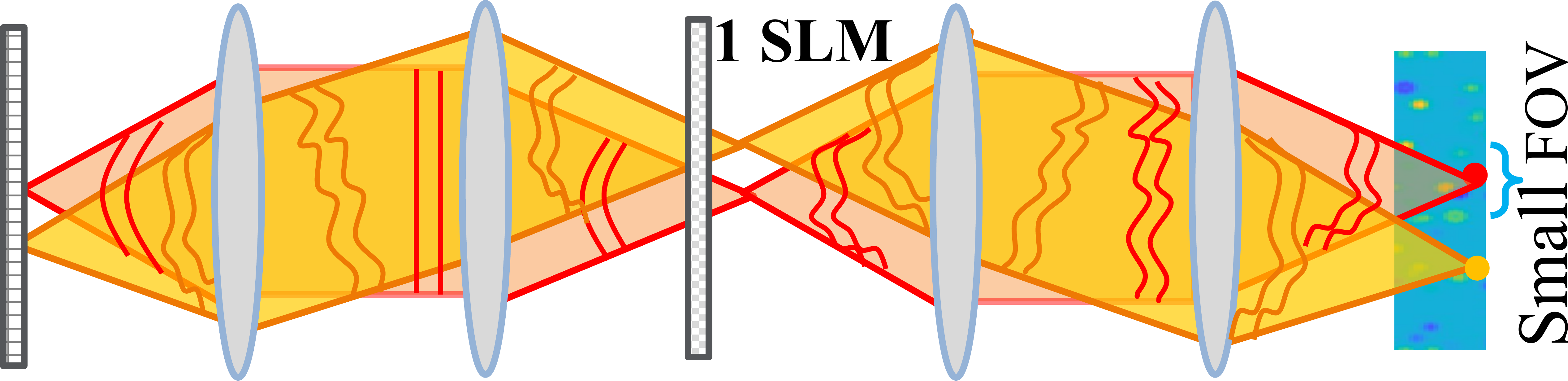}} \\[-0.7ex]
	\includegraphics[height=0.12\linewidth]{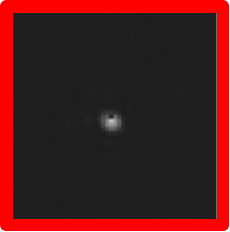} &                   \\
	\includegraphics[height=0.12\linewidth]{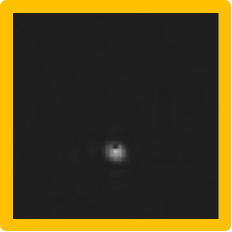} & \multirow{2}{*}[0.5cm]{\includegraphics[width=0.8\linewidth]{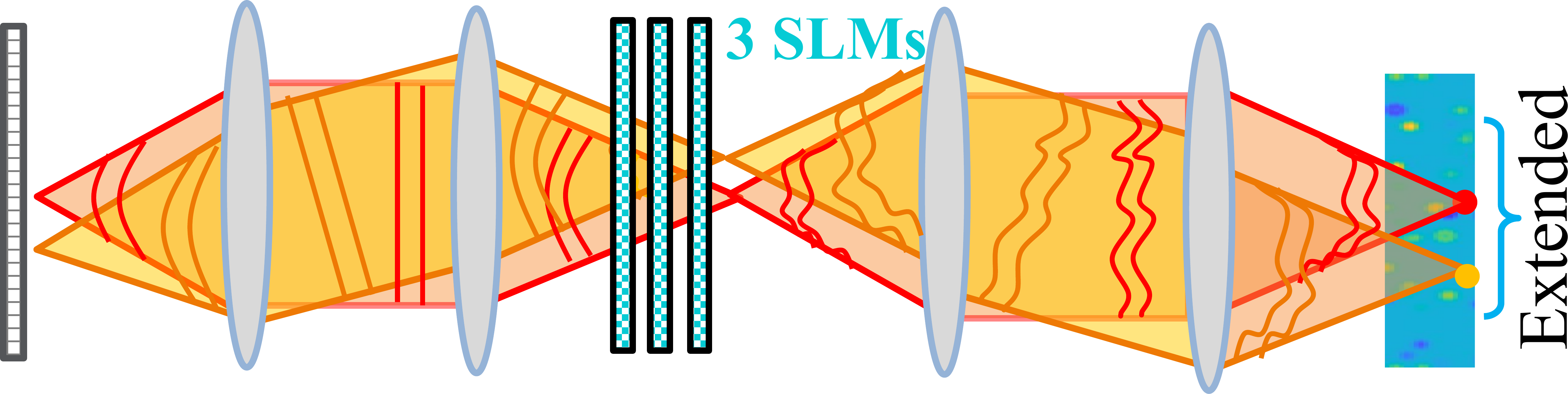}} \\[-0.7ex]
	\includegraphics[height=0.12\linewidth]{figs/setup/cam_focus_red.pdf} &                  
\end{tabular}
		\caption{{\footnotesize {\bf Layered wavefront shaping:} (Top) Consider a simple microscope setup where we use a 4F relay to image the light emerging from a tissue sample. Light  is aberrated on its way to the sensor, and we see speckle patterns rather than  sharp spots. (Middle) A single planar SLM can correct aberration for one spot (see red beam), but light  emerging from  nearby spots may not be corrected  (the yellow beam generates a speckle pattern on the sensor). (Bottom) The field of view of the correction can possibly be extended with a multi-layer correction. The goal is to form a correction replicating the 3D structure of volumetric aberration inside the tissue.
		The figure illustrates two wavefronts from nearby spots (red and yellow) passing through the same volumetric correction, and both focus into a sharp spot behind the tissue. } }\label{fig:setup-wavefront-shaping}\vspace{-1cm}
	\end{center}
\end{figure}

%% file: cvpr_sec_files_cr/fig_setup.tex
\begin{figure*}[!ht]
	\begin{center}
		\begin{tabular}{@{}c@{~~~~~~~~~}c@{~~~~~}c@{~~~~~}c@{}}	
			\includegraphics[height=0.12\textwidth]{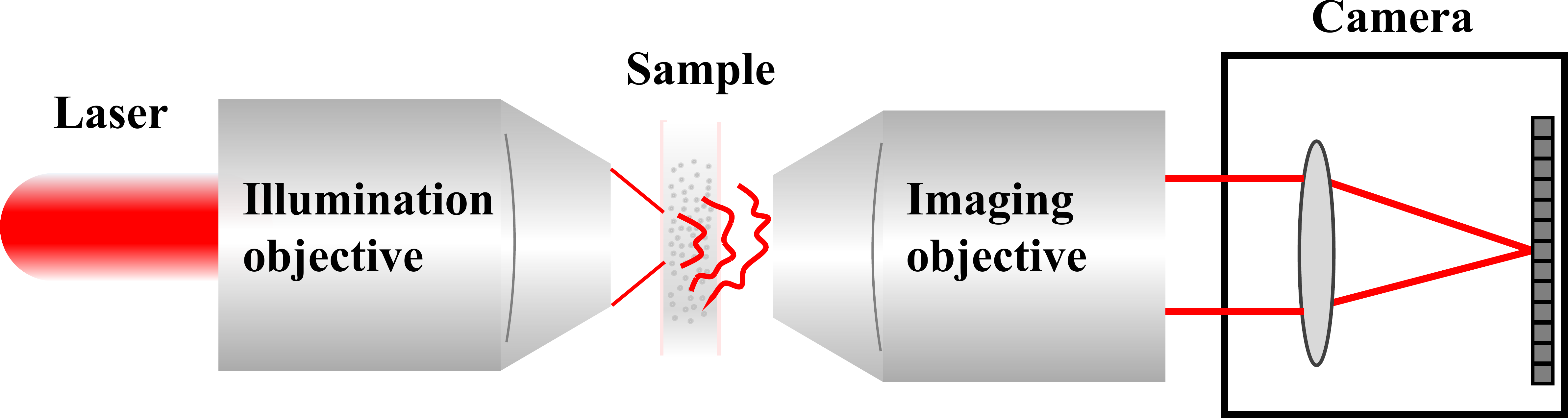}&
			\includegraphics[height=0.11\textwidth]{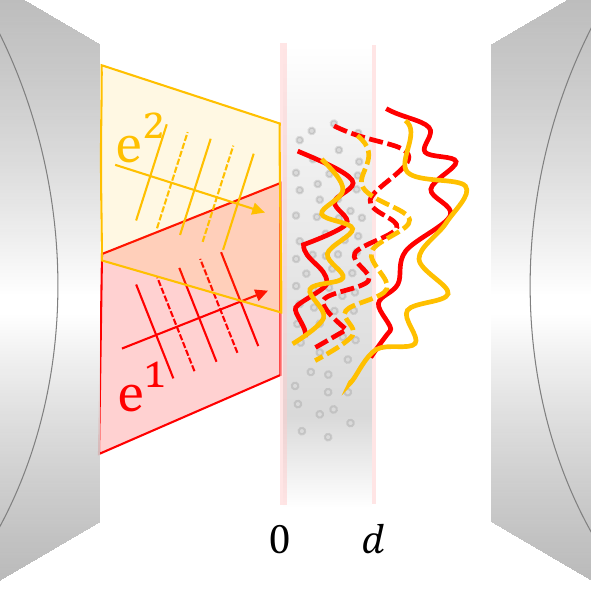}&
				\includegraphics[height=0.11\textwidth]{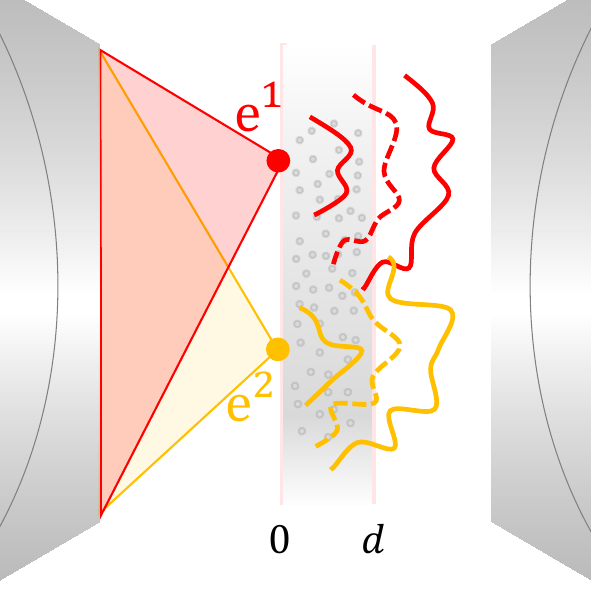}&	\includegraphics[height=0.11\textwidth]{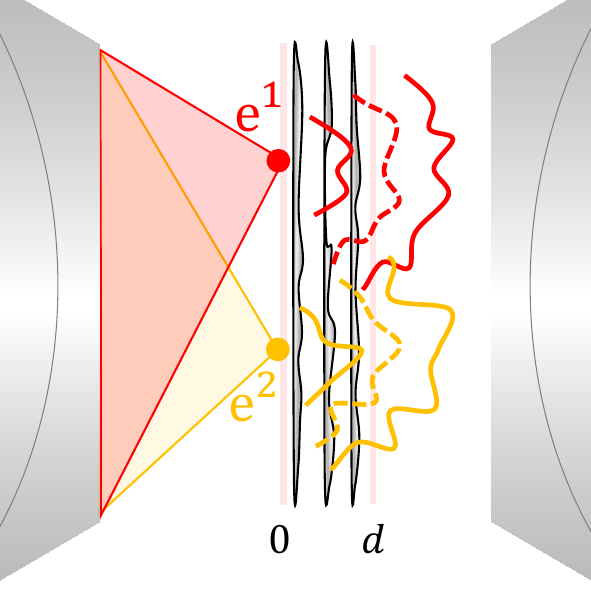}\\
			{\footnotesize(a) Full setup} &	{\footnotesize(b)  Fourier illum.} &{\footnotesize (c)  Point illum.} &{\footnotesize(d) Layered approx.}
		\end{tabular}\vspace{-0.2cm}
		\caption{\footnotesize {\bf Setup for transmission matrix acquisition:} (a) A setup for transmission matrix imaging where a set of basis illuminations  at the back of the sample  propagate through the volume, and the scattered wavefronts are measured by a camera at the other end. In (b,c) we zoom on the sample illustrating two possible illumination bases. (b) A set of directional wavefronts (Fourier basis). (c) A set of wavefronts focused to points at the edge of the volume (primal basis). Both transmission matrices can be related by a simple Fourier transform. (d) A multi-slice model approximates the volumetric scattering with a set of planar aberrations. }\label{fig:setup}

	\end{center}\vspace{-0.3cm}
\end{figure*}

%% file: cvpr_sec_files_cr/problem_formulation.tex

\vspace{-0.2cm}\section{Problem formulation} \vspace{-0.2cm}

\boldstart{Imaging setup:}
For  simplicity, the following derivation  considers coherent  scalar waves, ignoring polarization. 
In \figref{fig:setup}(a) we visualize a setup for transmission matrix acquisition.
A scattering medium, such as tissue, is illuminated by  a coherent  source $\be^k$ at one side of the medium. After propagating through the medium, the resulting coherent wavefront $\bt^k$ is recorded. Usually  $\bt^k$ are complex valued speckle patterns with a pseudo-random structure.  The illumination is translated and multiple  wavefronts are recorded and stored as columns of a transmission matrix $\TM$.

\boldstart{The transmission matrix:}
Since the propagation of coherent light is linear, if we want to predict the propagation of any wavefront $\bu$, we simply express it as a superposition of the recorded incoming illuminations $\bu=\sum_k \bu(k) \be^k$, where $\bu(k)$ are complex scalars, and  multiply it by the transmission matrix, to predict  $\TM\bu$.
The illumination wavefronts $\be^k$ can be selected at different bases. For example, the illumination can be translated at the Fourier plane, and then the wavefronts illuminating the sample are plane waves at different angles (\figref{fig:setup}(b)). Alternatively one can focus the light to a point at the back plane of the sample (\figref{fig:setup}(c)). This configuration is more appropriate for wavefront shaping algorithms whose goal is to recover wavefronts focusing inside the tissue and not far behind it. In both cases, however, the illuminations we can create are limited by the numerical aperture of the illuminating objective, which we mark here as $\NA$. Similarly, the scattered wavefronts are measured via an imaging objective with the same bounded $\NA$.  This $\NA$ would serve as a central component in the results derived below.  
For simplicity of resulting equations, for the rest of this manuscript we assume the imaging and illumination objectives are focused at the same plane, such that if we would remove the scattering sample and image the sources directly, we would see the $\be^k$ illuminations as sharp spots on the sensor plane. That is, if we use directional illuminations, we also measure the Fourier transform of the scattered waves, namely place the sensor at the Fourier plane of the imaging objective. If we use point sources at the illumination end of the medium, we also focus the imaging objective at the same plane, so we measure a speckle image of a relatively small support. 

\boldstart{Range and resolution:}
We assume that in the primal basis the illuminators are spaced within an area of $\prng\times \prng$, and we sample them at the Nyquist resolution, so for an illumination wavelength $\lambda$ they are separated by a pitch $\pstp=\lambda/(2\NA)$. Equivalently,  in the Fourier domain, we illuminate the sample with plane waves
of the form \vspace{-0.1cm}
{\small\BE \be(\ptd)=e^{{2\pi i}(\bunitkv\cdot \ptd)}\vspace{-0.1cm} \EE}where $\ptd$ is a point inside the 3D volume, and $\bunitkv=\frac1\lambda \dirvect $ where $\lambda$ is the wavelength and $\dirvect$ is a unit norm direction vector,  i.e.  the last entry of the direction vector $\omega_z$ is {\small$\omega_z=\sqrt{1-(\omega_x^2+\omega_y^2)}$.}  For simplicity of notation, we assume the refractive index (RI) of the leading medium is 1.
We can only use illumination direction inside the numerical aperture {\small$\sqrt{\omega_x^2+\omega_y^2}\leq \NA$} (or equivalently {\small$\sqrt{\xkv^2+\ykv^2}\leq \frng$} with {\small $\frng=\NA/\lambda$}).  
To sample the transmission matrix at the Nyquist pitch, the $\xkv,\ykv$ components of the illumination are sampled by the Nyquist pitch $\fstp=1/{2\prng}$.


With the above sampling scheme, the transmission matrices at the Fourier and primal bases are related by a simple Fourier transform $\TMp=\Fr^T\TMf \Fr$. While for wavefront shaping we are interested in primal transmission matrices, the Fourier ones simplify analysis.  Importantly, since Fourier and primal transmission matrices are related by an orthonormal transformation,  the fitting errors in both bases  are equivalent.


\input{cvpr_sec_files_cr/fig_small_full_zx_reconst}
\boldstart{The multi-slice model:}
The relation between an incoming illumination and an outgoing one is determined by the solution to a wave equation which takes into account the variation in refractive index (RI) inside the sample.
Rather than a fully continuous model, in a multi-slice model we approximate the volume with a set of planar aberrations, separated by small $z$ distances $\eps$, where we have free space propagation between layers, as in \figref{fig:setup}(d). Mathematically this can be described as \vspace{-0.1cm}
{\small\BE\label{eq:multi-layer-transmission-matrix}\vspace{-0.2cm}
\TMml=\OpPr_{O} \DiagM(\AbrVect_M) \ldots \OpPr_\eps \DiagM(\AbrVect_2)  \OpPr_\eps \DiagM(\AbrVect_1) ,
\EE}
where $\AbrVect_m(x,y)$ is the 2D aberration of the $m$th layer and $\DiagM(\AbrVect_m)$ is a diagonal matrix performing element-wise multiplication of the incoming wavefront by the aberration.   $\OpPr_\eps$ describes free space propagation between layers at distance $\eps$, which is essentially a convolution with a spherical wavefront. Finally $\OpPr_{O}$ propagates the wavefront in free space from the tissue boundary, via the lenses,  to the camera sensor. 

{\em Our goal  is to understand what plane spacing $\eps$ will facilitate a good representation of $\TM$, and how this approximation would degrade when we use sparser sampling.}

A common approximation for weak aberrations is:\vspace{-0.2cm}  
{\small\BE 
\AbrVect_m(x,y)=\mu+\dAbrVect_m(x,y),
\EE
}where $\mu$ is a scalar whose magnitude is a bit smaller than $1$, encoding the ballistic portion of the light which passes through the layer without aberration, and the magnitude of the aberration is significantly smaller than the ballistic term $|\dAbrVect_m(x,y)|\ll |\mu|$. 
For weak scattering most of the energy does not scatter more than once. It is therefore common to consider a first order approximation to the transmission matrix (often known as the single scattering, first-Born or Rytov approximation~\cite{Devaney:81,Lin:92,Chen:98,Nguyen2017,PMID:17694065}), which reduces to \vspace{-0.1cm}
{\small\BEA\label{eq:multi-layer-transmission-matrix-ss}
\!\!\TMwml\!\!\!\!&\!\!\approx\!\!& \!\!\!\mu^M\OpPr_{O} \OpPr_{M\eps}+\\ &&\!\!\!\mu^{M-1}\sum_{m=1}^M  \OpPr_{O} \OpPr_{(M-m)\eps} \DiagM(\dAbrVect_m) \OpPr_{m\eps}
\nonumber \vspace{-0.2cm} \EEA}
The full light propagation in \equref{eq:multi-layer-transmission-matrix} includes multiplication of the different aberrations and hence the transmission matrix is highly nonlinear.   In contrast, the advantage of the weakly scattering model in \equref{eq:multi-layer-transmission-matrix-ss} is that the resulting transmission matrix is a  linear function of the aberration layers. This linear approximation significantly simplifies analysis.

Diffraction tomography algorithms such as~\cite{Chowdhury:19} capture a subset of columns from the transmission matrix and use gradient descent optimization to seek layers $\AbrVect_1,\ldots,\AbrVect_M$ which best explain the measured wavefronts, assuming that the propagation operators $\OpPr_\eps,\OpPr_{O}$ are known. We will use a similar optimization scheme below, when trying to find a multi-slice representation for a given transmission matrix. 
To correct aberrations, a physical multi-conjugate wavefront shaping system would display on the SLM the conjugate of the recovered layers.  

%% file: cvpr_sec_files_cr/fig_small_full_zx_reconst.tex
\begin{figure*}[t!]
	\begin{center}
		\begin{tabular}{@{}c@{}}	
			\includegraphics[height=0.3\textwidth]{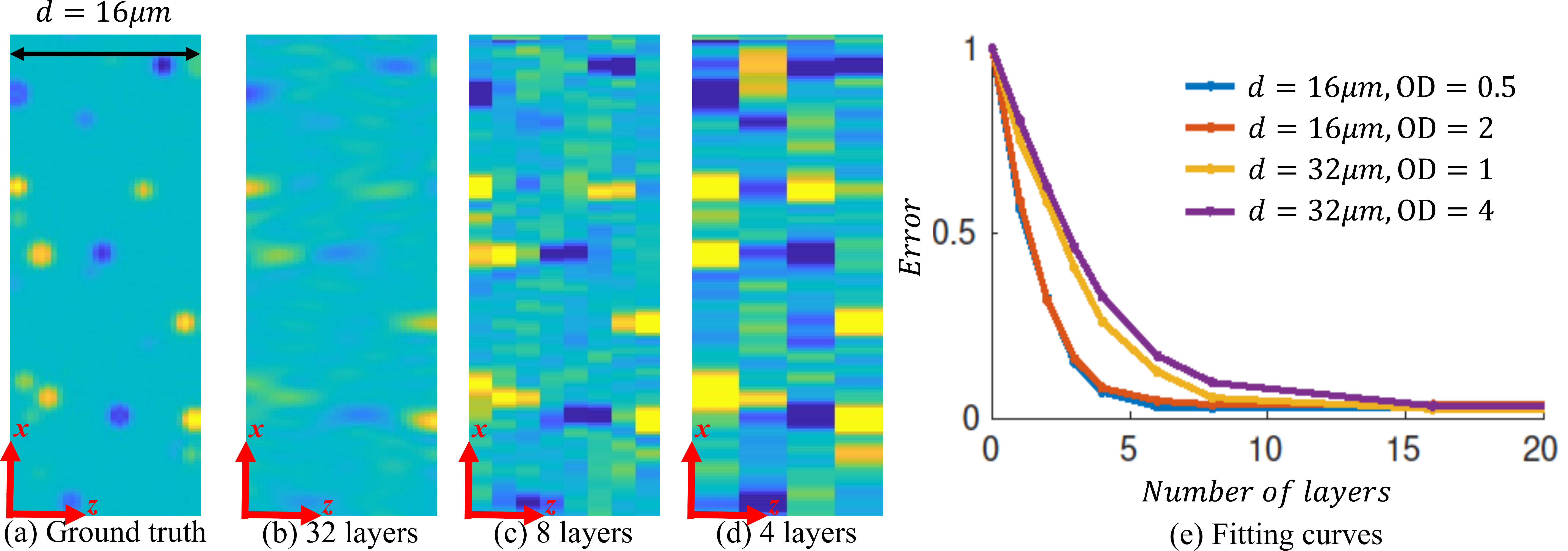}
		\end{tabular}\vspace{-0.2cm}
		\caption{ \footnotesize {\bf Reconstructed volumes:} (a) An xz slice from a toy volume, visualizing its refractive index variations. (b-d) Reconstructions with varying layer numbers. The columns of the transmission matrix provide an under-constraint inversion problem, and even when a dense set of layers is allocated in b, the axial resolution of the reconstructed volume is low. As a result the sparse reconstructions in c,d can also well explain the transmission matrix. (e) Numerical evaluation of fitting error as a function of the number of layers. Transmission matrices can be well fitted even if the allocated number of layers is lower than Nyquist predictions. However the required number of layers increases with the tissue thickness. The layer number is less affected when the optical density is increased for the same tissue thickness.
			}\label{fig:full_xz_reconst}
	\end{center}\vspace{-0.2cm}
\end{figure*}

%% file: cvpr_sec_files_cr/fittingevals.tex
 \input{cvpr_sec_files_cr/fig_small_sup_zx_reconst}
\section{Evaluating multi-slice approximations}
We start with numerical experiments demonstrating the main findings. In \secref{sec:samplingtheory} below we derive a theory explaining them.
To gain intuition we generated a toy volume composed of spheres with varying refractive indices illustrated in \figref{fig:full_xz_reconst}(a), and obtained the corresponding transmission matrix by propagating a set of point sources through the volume. We then used the transmission matrix to solve for a multi-slice approximation, while varying the number of layers. In \figref{fig:full_xz_reconst}(b-d) we visualize some of the volumetric reconstructions, assuming a $\lambda=500 nm$ illumination.
Even when the fitted layers are sampled very densely at an interval of $\eps=0.5\mu m$, we cannot fully recover the target and the fitted volume has a low axial resolution. We analyze this in \secref{sec:samplingtheory} below and explain that this results from the well known missing cone problem, a fundamental degeneracy limiting the axial resolution of a microscope. Said differently the transmission matrix, or the full set of propagating waves, only provides a set of  under-constraint measurements from which the volume cannot be uniquely recovered. While the missing cone problem is considered a fundamental limitation in microscopy, we argue that in the context of wavefront shaping it actually turns into an advantage. Since the volume reconstruction is not unique we  can approximate the same transmission matrix with a sparser set of layers.

To quantify the fitting error of the multi-slice model, we denote by $\errTap_M$ the reconstruction error of a given transmission matrix with the best $M$ layers
\vspace{-0.2cm}{\small \BE\label{eq:err-TM-M-layers}\errTap_M=\min_{\AbrVect_1,\ldots, \AbrVect_M}\|\TMcap-\TMml\|^2.\vspace{-0.1cm}\EE}
In \figref{fig:full_xz_reconst}(e) we evaluate $\errTap_M$ as a function of the layer number $M$.  Even the seemingly aliased fit of \figref{fig:full_xz_reconst}(d) still provides a low fitting error. 
However, there are no free lunches and when the number of layers is too low the fitting error is high.
 When we repeat the simulation with a thicker volume we need even more layers for a good fit.
 Interestingly, we also repeat the simulation with denser volumes (larger optical depth, or shorter mean free path) of the same thickness, and this has less effect on the required number of layers. 
 In our plots, the case $M=0$ refers to no correction at all, so effectively we assume that the transmission matrix is diagonal. The similarity between a diagonal transmission matrix and the target transmission matrix is a measure of the amount of ballistic light.

 The thickness of the synthetic simulation here is limited due to computational constraints. However, usually wavefront shaping systems attempt to see into a few hundred microns of tissue and we anticipate that for such volumes we need at least a few dozen layers for a good fit. Clearly realizing a physical wavefront shaping system with that many layers is impractical. 
 We argue that while a small number of layers is not enough for focusing light anywhere across the tissue area, it can still help us to focus light over a compact support area, and this area is significantly wider than what we can correct with a single layer.

 \boldstart{Compact support:}
 To test the effect of the target area, we consider transmission matrices measured with illumination points confined to a limited area of size  $\prng\times\prng$. If we can fit such a transmission matrix with $M$ layers and place the values of these $M$ layers on $M$ SLMs, the resulting wavefront shaping system can correct light from the  $\prng\times\prng$ area, but target points outside this area are unlikely to be corrected as they are not included in the fitting process.
 
  To test the effect of support on the number of fitted layers, we return to 
 the numerical simulation of \figref{fig:full_xz_reconst}, but fit  subsets of columns contained in smaller $\prng$ ranges.   Indeed, smaller supports can be well-fitted with a smaller number of layers.
 As another way to understand it, we show in \figref{fig:support_xz_reconst} an xz slice from the ground truth refractive index volume and from a few reconstructions. We show that even if we do not limit the layer number $M$, the recovered RI volume has limited axial resolution and this axial resolution is lower when measured range $\prng$ decrease.
  This explains why lower supports can be fitted with fewer layers. For each of the two supports we also show a sparse fit,
  with the minimal  $M$ value that provides a good prediction of the transmission matrix.
  We explain this observation in \secref{sec:samplingtheory}.
  This result is also in agreement with
  ~\cite{ChoiLyers2023}, who correct aberration over a relatively {\em small field of view} through a few hundred microns of tissue, using only 5 layers. 
  
  \input{cvpr_sec_files_cr/fig_fit_chicken_varysup}

  \boldstart{Implications to layered wavefront shaping systems.}
  In \figref{fig:support_xz_reconst}(f) we plotted a bar below which the approximation is reasonable. We can see that for e.g. the $40\times 40\mu m$ support we can obtain good focusing with about $M=6$ layers. It is likely that fitting a full field-of-view of a few hundred microns would require more layers.   However, we can see that while with $M=1$ layer we could only focus over a small area of about $1\times 1\mu m$, when increasing to $M=2$ we could focus over a $\times 4 $ larger area of support $2\times 2\mu m$, and with  $M=3$ we could focus over a $\times25$ larger area of support $5\times 5\mu m$. Computing multiple correction elements is only going to increase the complexity of the wavefront shaping algorithm in~\cite{DrorNatureComm24}  linearly with the number of layers. 
  Alternatively, a wavefront shaping system using  a single SLM would need to scan the field-of-view sequentially  and correct each local window independently. Since the increase in the support of the window we can explain with e.g. $3$ layers is significantly larger than a $\times3$ factor, a multi-conjugate wavefront shaping system with a small number of SLMs as in~\cite{Monin2022Cascade,Thaung:09,Laslandes:17,Furieri23} can already have a large practical benefit.
  While \figref{fig:support_xz_reconst} only considers a simulation, we analyze realistic numbers below. 

%% file: cvpr_sec_files_cr/fig_small_sup_zx_reconst.tex
\begin{figure*}[t!]
	\begin{center}
		\begin{tabular}{@{}c@{}}	
			\includegraphics[height=0.3\textwidth]{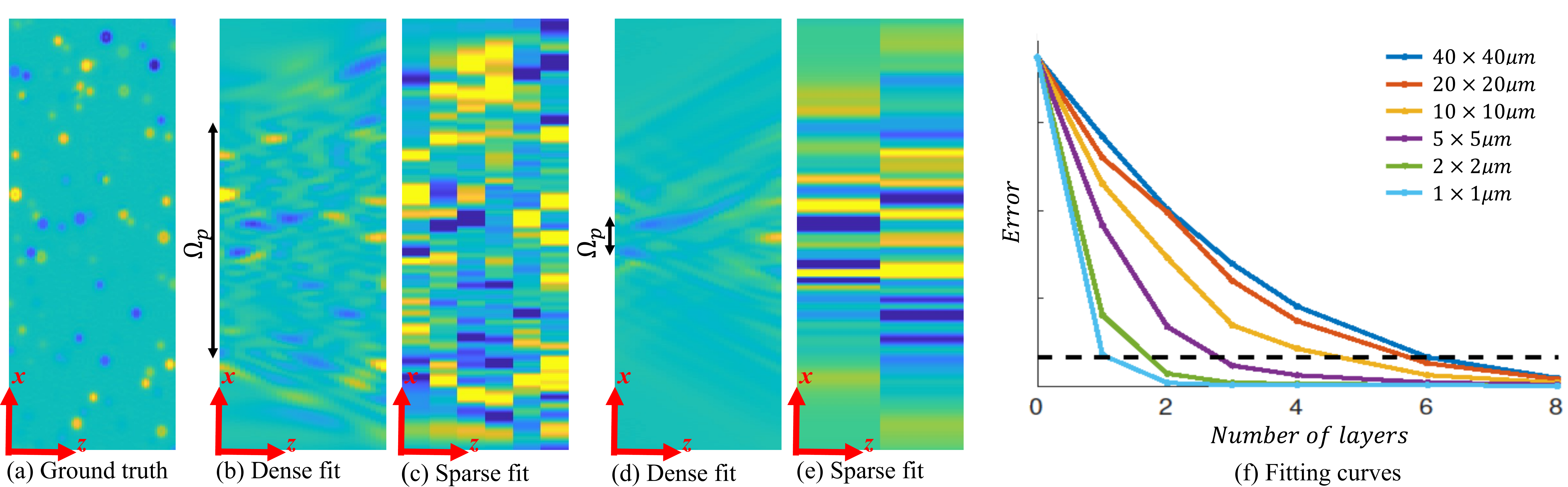}
		\end{tabular}\vspace{-0.2cm}
		\caption{ \footnotesize {\bf Reconstructed volumes with varying supports:} (a) An xz slice from a toy volume, visualizing its refractive index variations. (b,d) Two reconstructions of the volume using a dense layer sampling. When the support of the measured window $\prng$ shrinks, the lateral extent of the reconstructed region decreases, however, its axial resolution is also reduced. (c,e) A sparse fit with the same input. Despite the low quality of these reconstructions, they predict well the columns of the transmission matrix of the given support $\prng$.
			(f) Numerical evaluation of fitting error as a function of the number of layers,  while varying the range $\prng$ covered by the transmission matrix.   Smaller supports can be fitted with fewer layers.}\label{fig:support_xz_reconst}
	\end{center}\vspace{-0.2cm}
\end{figure*}

%% file: cvpr_sec_files_cr/fig_fit_chicken_varysup.tex
\begin{figure*}[!ht]
	\begin{center}
		\begin{tabular}{@{}c@{~~~~~~~}c@{}}	
		
				\begin{tabular}{@{}c@{}}		
				\includegraphics[height=0.24\textwidth]{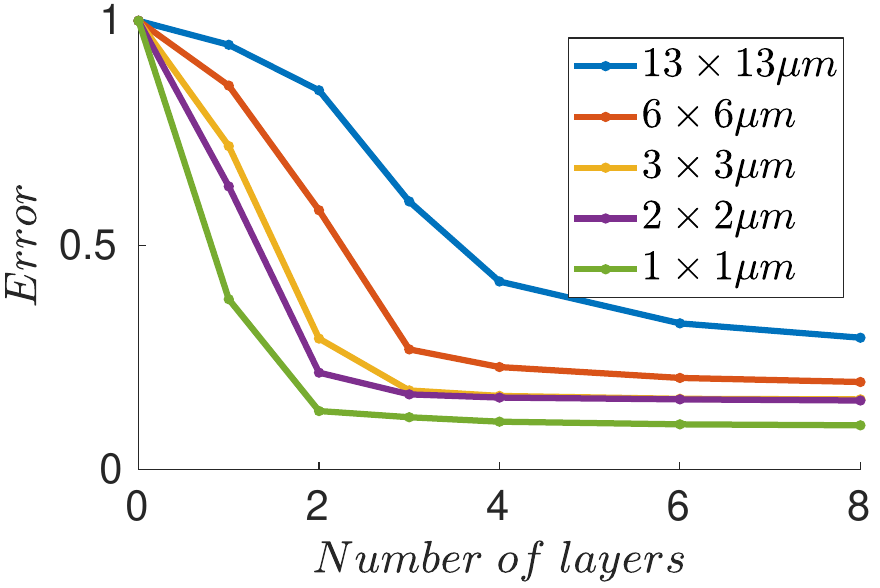}
			\end{tabular}&
			\begin{tabular}{@{}c@{}c@{}c@{}@{}c@{}c@{}c@{}@{}c@{}c@{}c@{}@{}c@{}c@{}c@{}c@{}c@{}}	
				{\raisebox{0.60cm}{\rotatebox[origin=c]{90}{$13\times 13\mu m$ }}}&
				\includegraphics[height=0.08\textwidth]{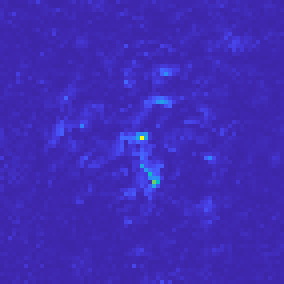}&
				\includegraphics[height=0.08\textwidth]{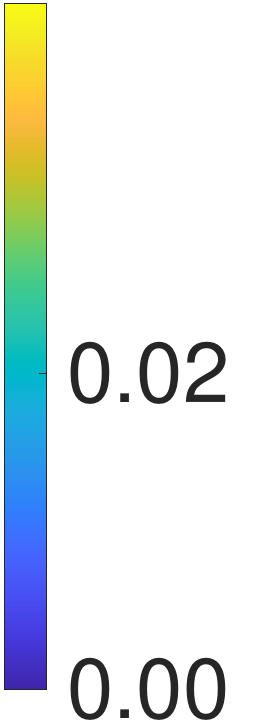}&		
               \includegraphics[height=0.08\textwidth]{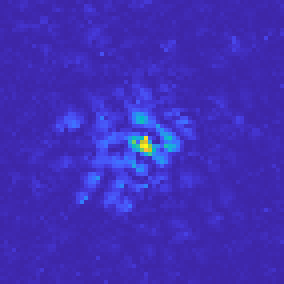}&
               \includegraphics[height=0.08\textwidth]{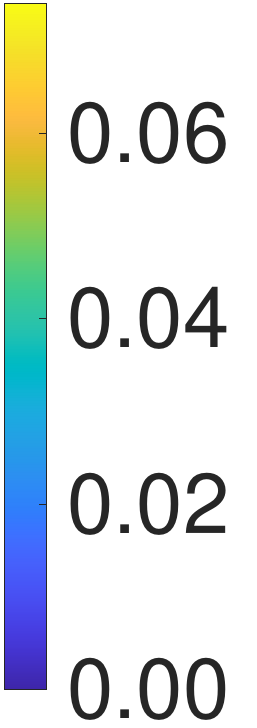}&		
               \includegraphics[height=0.08\textwidth]{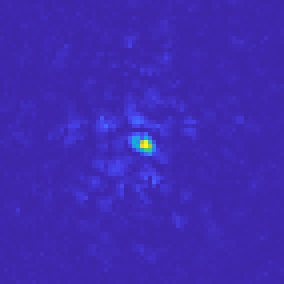}&
               \includegraphics[height=0.08\textwidth]{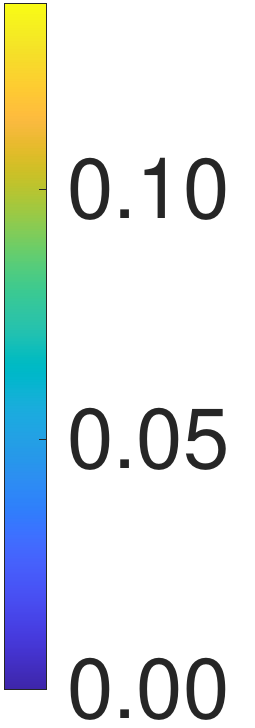}&		
               \includegraphics[height=0.08\textwidth]{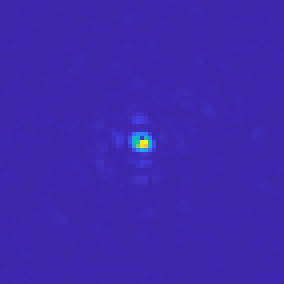}&
               \includegraphics[height=0.08\textwidth]{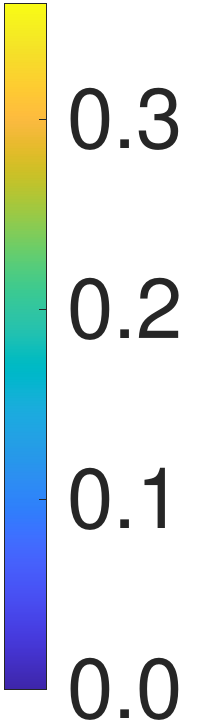}&				
               \includegraphics[height=0.08\textwidth]{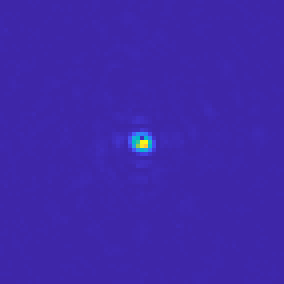}&
               \includegraphics[height=0.08\textwidth]{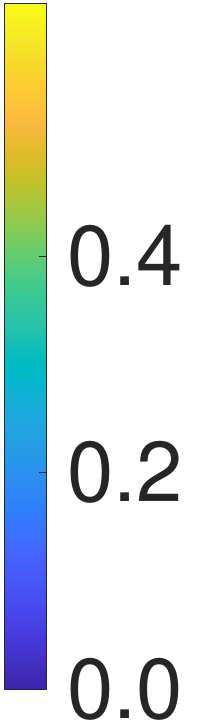}
               \\
               		{\raisebox{0.60cm}{\rotatebox[origin=c]{90}{$6\times 6\mu m$ }}}&
               	\includegraphics[height=0.08\textwidth]{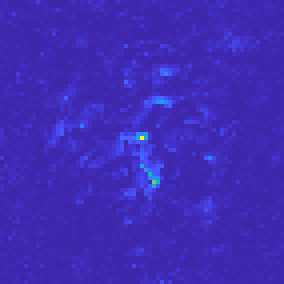}&
               \includegraphics[height=0.08\textwidth]{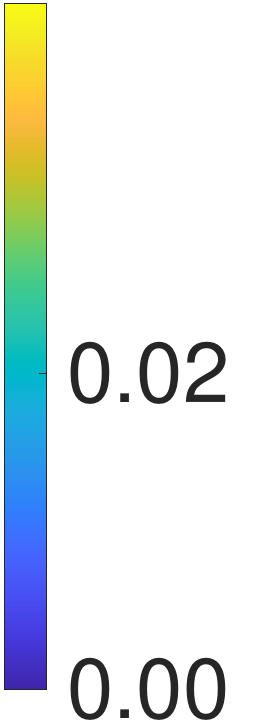}&		
               \includegraphics[height=0.08\textwidth]{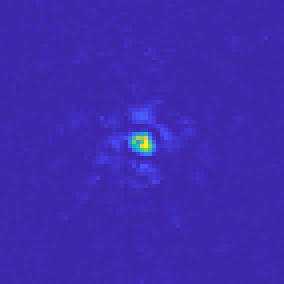}&
               \includegraphics[height=0.08\textwidth]{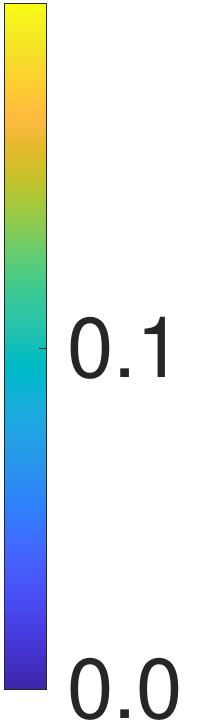}&		
               \includegraphics[height=0.08\textwidth]{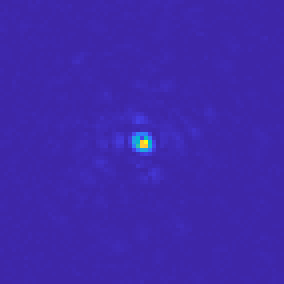}&
               \includegraphics[height=0.08\textwidth]{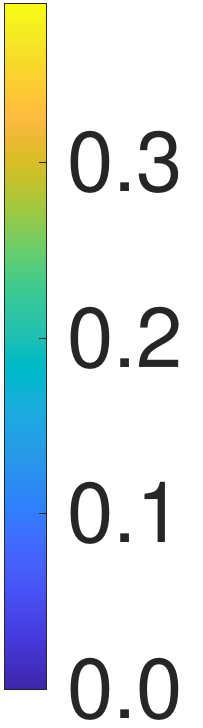}&		
               \includegraphics[height=0.08\textwidth]{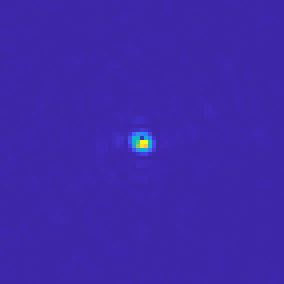}&
               \includegraphics[height=0.08\textwidth]{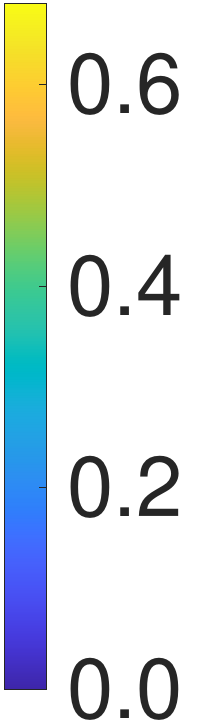}&				
               \includegraphics[height=0.08\textwidth]{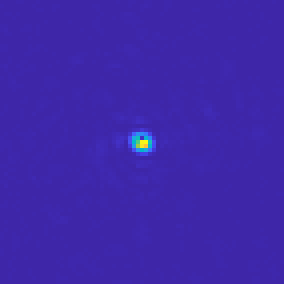}&
               \includegraphics[height=0.08\textwidth]{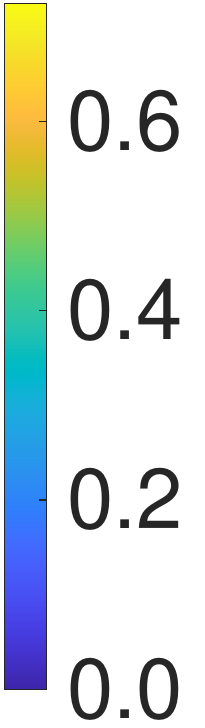}
               \\
               	{\raisebox{0.60cm}{\rotatebox[origin=c]{90}{$3\times 3\mu m$ }}}&
               	\includegraphics[height=0.08\textwidth]{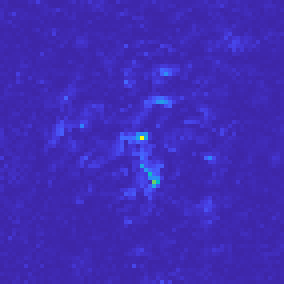}&
               \includegraphics[height=0.08\textwidth]{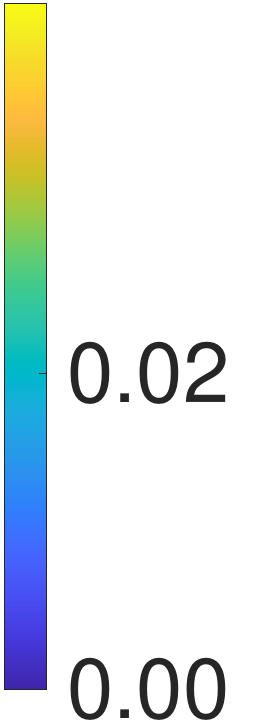}&		
               \includegraphics[height=0.08\textwidth]{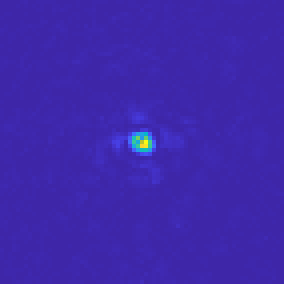}&
               \includegraphics[height=0.08\textwidth]{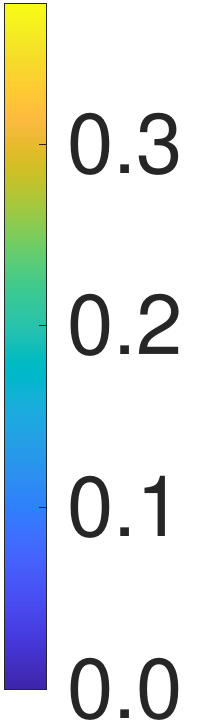}&		
               \includegraphics[height=0.08\textwidth]{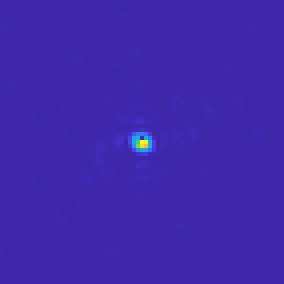}&
               \includegraphics[height=0.08\textwidth]{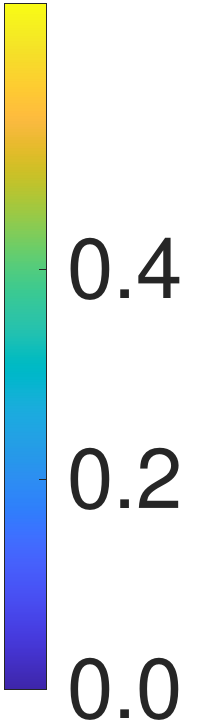}&		
               \includegraphics[height=0.08\textwidth]{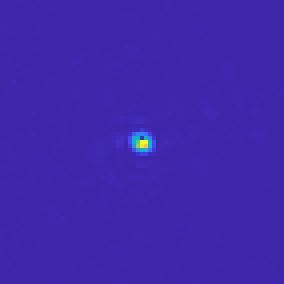}&
               \includegraphics[height=0.08\textwidth]{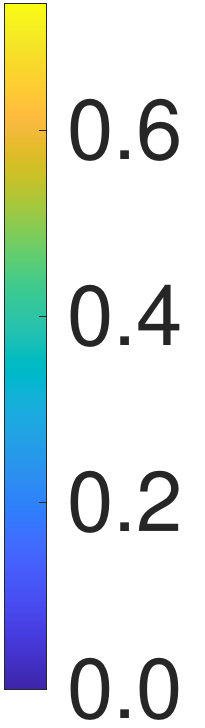}&			
               \includegraphics[height=0.08\textwidth]{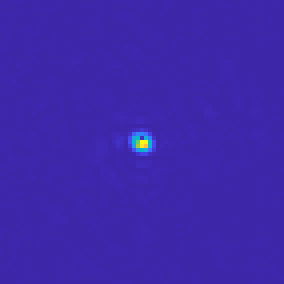}&
               \includegraphics[height=0.08\textwidth]{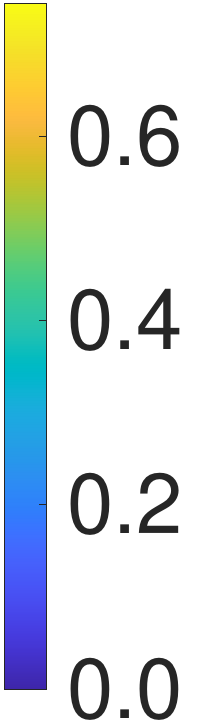}
               \\
               &$M=0$&&$M=1$&&$M=2$&&$M=3$&&$M=6$
			\end{tabular}
		\end{tabular}\vspace{-0.2cm}

\caption{\footnotesize {\bf Layer support for a captured transmission matrix:} We consider a transmission matrix captured in the lab through a chicken breast layer of thickness $170\mu m$. Left:  plotting the fitting error as a function of the number of layers.   Smaller supports can be fitted with fewer layers.    Right: Visualization of a spot focusing behind the tissue computed using the wavefronts of the fitted model.  We compare models fitted to different supports. In the top row we need a larger number of layers, but the spot can be scanned over a $25\times 25\mu m$ window using the same layers. In the lower rows we achieve a focused spot using a smaller number of layers, but these layers can scan the focused spot over smaller windows of sizes $6\times 6\mu m$ and $1\times 1\mu m$.	}\vspace{-0.3cm}

	\label{fig:fitting-chicken-varysup}
	\end{center}
\end{figure*}

%% file: cvpr_sec_files_cr/empirical_fits.tex

\boldstart{Lab results:}
In the supplementary file we validate the theory with multiple real transmission matrices, representing thick volumes with multiple-scattering.

We start with numerically simulated transmission matrices using a more accurate wave-propagation model \cite{Bar:2019,Bar:2020}.
We then test lab-captured transmission matrices measuring realistic scattering samples including mouse brain tissue, chicken breast tissue and a parafilm layer. {\em Please refer to supplementary file for additional results}.

In \figref{fig:fitting-chicken-varysup} we show one fitting result using a multiply-scattering chicken breast layer of thickness 
 $\smpthick=170\mu m$. The transmission matrices cover an area with support $\prng\times \prng=13\times 13\mu m$. 
 As in \figref{fig:support_xz_reconst}(f), we see that smaller spatial ranges can  be fitted with fewer layers. 

 As another way to asses the quality of the fit, we used columns of the fitted matrices to focus light through the tissue and show the shape of the focused spot and visualize the focusing quality in \figref{fig:fitting-chicken-varysup}. To simulate the focused spot we simply  visualize the product $\TMcap \cdot \bt^k_{\text{fit,M}}$, where $\bt^k_{\text{fit,M}}$ is a column of the fitted matrix $\TMml$, namely a wavefront generated by the layered approximation.
 Note that while we only show focusing at one point, the layers are optimized such that they allow us to focus at {\em any point} inside the $\prng\times \prng$ window. 
 
 From the results in \figref{fig:fitting-chicken-varysup} we can see that for the $13\times 13\mu m$ support measured, we can obtain good focusing with about $3$ layers. 	Fitting a full field-of-view of several hundred microns likely requires additional layers. However, we observe that while using $M=1$ allows us to focus on a small area of about $3\times 3\mu m$,  with $M=3$, we can focus on an area $ \times 18$ larger, reaching  $13\times 13\mu m$. Therefore, using multiple layers can significantly accelerates a sequential scanning of a wave-front shaping system.

%% file: cvpr_sec_files_cr/missingcone.tex
\vspace{-0.1cm}\section{Sampling theory for transmission matrices}\vspace{-0.1cm}\label{sec:samplingtheory}
Our goal in this section is to explain the empirical findings of the previous section by analyzing  the information captured by a transmission matrix.
\subsection{Nyquist sampling and the missing cone}\vspace{-0.1cm}
We start by considering the case of weakly scattering volumes which, as in \equref{eq:multi-layer-transmission-matrix-ss}, can be reasonably approximated by considering the component of light that scattered once. We also assume the transmission matrix is sampled over a very wide support $\prng$.
 We discuss the implications of multiple scattering and  limited supports  in subsequent sections.  

We denote  by $\refin(\ptd)$ the difference between the RI at a 3D position $\ptd$ in the volume to that of the leading medium around it.  We denote its 3D Fourier transform by $\frefin(\bkv)$. 

\input{cvpr_sec_files_cr/fig_cone_parts}
We start by considering the Fourier representation of the transmission matrix, where it is illuminated and measured by plane waves. Using the weakly scattering (first-Born) approximation,   the wavefront scattering toward direction $\dirvect^v$ when illuminated by an incoming plane wave at direction $\dirvect^i$ is proportional, up to a multiplicative factor, to \cite{WOLF1969153,mertz2019introduction,Lauer2002}:
\vspace{-0.1cm}{\small \BE\label{eq:single-scat-tm}
\propto \int \refin(\ptd) e^{{2\pi i}{}{(\bunitkv^i-\bunitkv^v)\cdot \ptd}} d\ptd, \vspace{-0.1cm} \EE
}with $\bunitkv^v=1/\lambda\dirvect^v$, $\bunitkv^i=1/\lambda\dirvect^i$.
This implies that effectively an entry of the transmission matrix is a sample from the 3D Fourier transform of the volume at frequency $\bkv=\bunitkv^i-\bunitkv^v$. Note that throughout this paper we use bars $\dirvect, \bunitkv$ to denote vectors with norm constraints, and use normal bold  fonts to denote standard 3D vectors, such as frequencies $\bkv$ in the 3D spectrum. 
Unfortunately, a large portion of the 3D spectrum cannot be measured.   Recalling that the illumination and viewing directions are limited by an aperture of width $\NA$, the subset of 3D frequencies captured by the transmission matrix occupies only a butterfly shape, and the rest of the cone is missing, see illustration in \figref{fig:cone-parts}(a).
Using paraxial approximation it can be shown~\cite{WOLF1969153,Lauer2002} that the range of frequencies which can be generated is 
\vspace{-0.1cm}
\BE\label{eq:butterfly-shape}
{\small|\zkv|\leq \NA \xykvnor -\frac{\lambda}{2}\xykvnor^2, \quad {\text{with}}\quad \xykvnor=\sqrt{\xkv^2+\ykv^2}.}
\vspace{-0.1cm}\EE

 The missing cone is a serious problem for diffraction tomography algorithms aiming to recover the 3D structure of the refractive index $\refin(\ptd)$ from columns of the transmission matrix, since many of the frequencies of the volume cannot be recovered.  We argue that in the context of transmission matrices, the missing cone problem turns into an advantage. The fact that the transmission matrix only captures a limited range of frequencies allows us to approximate it with fewer samples, as derived below.

\boldstart{Minimal sampling:} Our goal is to approximate a transmission matrix with layers. We start by asking  what  spacing between layers will allow an exact reconstruction of the transmission matrix.
If we approximate the volume $\refin(\ptd)$ with a set of planes, we essentially sample a continuous signal. Classical Nyquist theory states that if we sample a signal at intervals of $\eps$, its Fourier transform includes replicas spaced by $\repspacing={1}/{\eps}$.
To achieve a good reconstruction, we want replicas to happen only in areas of the spectrum that the transmission matrix does not sample.
Since the $\xkv,\ykv$ components of our illuminations are limited by the aperture $\xykvnor\leq \frng$, and
given the butterfly shape in \equref{eq:butterfly-shape}, the  $\zkv$ frequencies we need to maintain are bounded to
\vspace{-0.1cm}{\small\BE
|\zkv|\leq  \frac{\NA\frng}{ 2}. \vspace{-0.1cm}
\EE}
Therefore, for an error-free reconstruction, the minimal Fourier range and primal spacing are
\vspace{-0.2cm}{\small \BE\label{eq:opt-sampling}
\repspacing^*=\NA\frng,\quad \eps^*=\frac{1}{\NA\frng}=\frac{\lambda}{\NA^2}.\vspace{-0.1cm}
\EE } For a scattering volume of thickness $\smpthick$, the number of required layers is: 
\vspace{-0.1cm}{\small \BE\label{eq:Mopt-def}
M^*=\frac{d \cdot \NA^2}{\lambda}.\vspace{-0.1cm}
\EE}
For typical numbers, consider e.g. $\lambda=0.5\mu m$ and $\NA=0.5$, the plane spacing should be  relatively small, and $\eps\leq 2\mu m$. For a modest tissue  thickness of $d=200\mu m$, we will need $M=100$ planes. Clearly it is impractical to realize such a large number of planes by a physical correction system.
\vspace{-0.1cm}
\subsection{Sparse multi-slice approximations}\vspace{-0.1cm}

Consider a scattering sample of thickness $\smpthick$, which we try to approximate using $M<M^*$ layers separated by $\eps=\frac{\smpthick}{M}$.
Our goal is to show that the fitting error $\errTap_M$ defined in \equref{eq:err-TM-M-layers} decays relatively fast with $M$ and we can get a reasonable approximation to the transmission matrix even if the number of layers is lower than the exact prediction in \equref{eq:Mopt-def}.
This is a result of two main properties: (i)~The volumes describing realistic tissue samples have more energy in the low frequencies; and (ii)~the structure of the missing cone implies that, in any case, a significant portion of the spectrum is not sampled by the transmission matrix.

To understand this, consider a naive selection of $M$ layers $\AbrVect_1,\ldots, \AbrVect_M$. Rather than actually solving an optimization problem,  we use the ground truth volume $\frefin(\bkv)$, and simply set to zero any frequency content of the RI volume  $\frefin(\bkv)$ at $\zkv$ values larger than the possible Nyquist range \vspace{-0.2cm}{\small\BE\label{eq:Nq-band-smp}\repspacing^M=\frac{1}{\eps}=\frac{\lambda M}{\smpthick},\vspace{-0.1cm}\EE}and we then Fourier transform  $\frefin(\bkv)$ to $\refin(\ptd)$ and sample planes at spacing $\eps=d/M$. The error of this approximation is basically the integral of content above the cut-off $\repspacing^M$. 
Since according to \equref{eq:Nq-band-smp}, the cut-off frequency  $\repspacing^M$ scales linearly with the number of layers, the portion of the spectrum which is lost by this low-pass operation scales linearly with the number of layers $M$. Thus, the naive answer is that the error of a multi-slice approximation decays linearly with $M$.
In practice we show below that the error decays much faster, since 
 {\em large areas from the 3D spectrum of the sample are  not used by the transmission matrix. } 

To gain intuition,  consider \figref{fig:cone-parts}(b). 
We note that the butterfly shape is such that low 2D frequencies (i.e. low $\xykvnor$),  also span less content along the $\zkv$ axes, and hence these frequencies  are not lost even with low bandwidths $\repspacing^M$.
For the higher $\xykvnor$ frequencies, the lower $\zkv$ part, marked in dashed blue in \figref{fig:cone-parts}(b) is preserved, while the higher portion marked in red is lost.

Another important property of tissue is that its RI is locally smooth, and therefore if we look at its spectrum (\figref{fig:cone-parts}(c)), we have much more energy in low frequencies than in high ones. This also agrees with the fact that tissue is forward scattering. Therefore, in the red areas of \figref{fig:cone-parts}(b) that are not sampled by the transmission matrix, there is less energy than in the lower frequencies. Thus, despite the fact that the multi-slice approximation sacrifices such frequencies, not much energy is being lost.

In most cases, it is hard to give analytic formulas for the decay of $\errTap_M$ as a function of $M$, but the supplementary file contains some derivation under simplifying assumptions.


%
%

\boldstart{Multiple scattering.}
The results in \figref{fig:full_xz_reconst}(e) include multi-slice fits with both lower and  higher optical depth values up to $ \od=4$, showing that the fitting error still decays rapidly with the number of layers.

While it is hard to give analytical predictions for the multiple scattering case, it appears that when the optical depth of the tissue is not too high, similar results hold. The reason is that  for a forward-scattering material such as tissue, light paths  only scatter at small angles, and  after a small number of scattering events they are not completely scrambled. For an intuition, in the supplementary file we use a Monte-Carlo simulation illustrating  that for a forward-scattering material,  after a modest number of scattering events, most light paths still remain within the butterfly area of the spectrum.
This is usually the regime that wavefront-shaping algorithms like~\cite{Dror22,DrorNatureComm24} attempt to tackle, as such algorithms attempt to push the depth at which conventional microscopes can see, but they do not yet attempt to image extremely deep where scattering is fully diffused.

%% file: cvpr_sec_files_cr/fig_cone_parts.tex
\begin{figure}[!t]
	\begin{center}
		\begin{tabular}{@{}c@{~~}c@{~~}c@{}}	
			\includegraphics[height=0.15\textwidth]{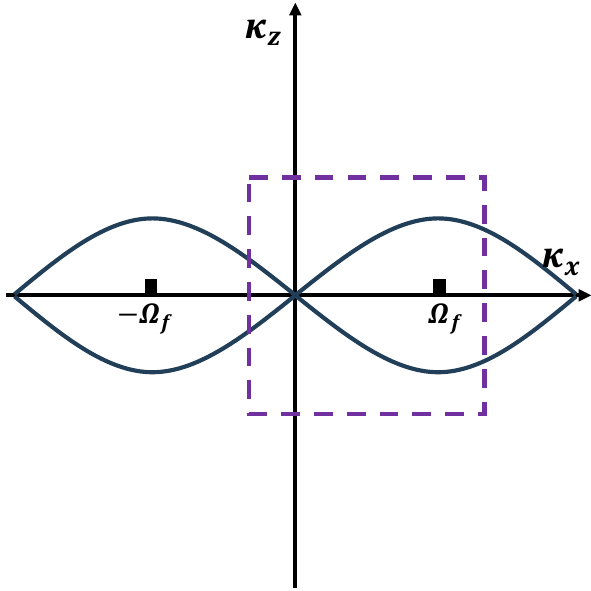}&
			\includegraphics[height=0.15\textwidth]{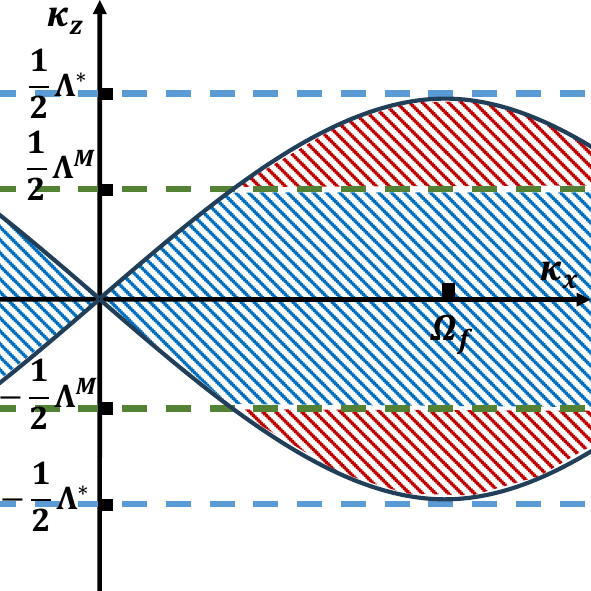}&
			\includegraphics[height=0.15\textwidth]{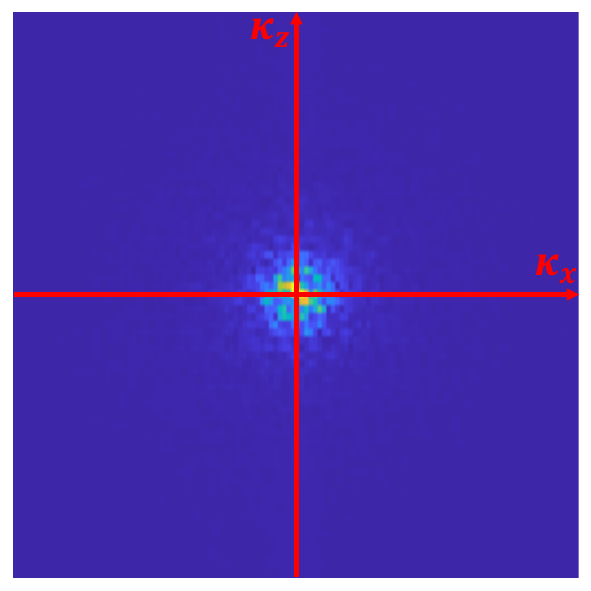}
			\\(a)&(b)&(c)
		\end{tabular}\vspace{-0.2cm}
		\caption{\footnotesize {\bf Spectrum structure:} (a) An $xz$ slice out of the spectrum of   $\frefin(\bomg)$. Entries of a weakly-scattering  transmission matrix limited by an aperture $\NA$ only lie inside the butterfly area. (b) Zooming on the center right area of (a) (purple square). Assuming the $\omega_z$ axis is cut at $\pm \frac12 \repspacing^M$, the transmission matrix entries inside the dashed blue area are maintained, and the entries in the dashed red area are lost. (c) The energy at a slice from the spectrum of the volume in \figref{fig:full_xz_reconst}(a). Since the volume structure is smooth, most energy is concentrated at the low frequencies. As a result the cropped red region in (b) do not contain a lot of energy.	}\label{fig:cone-parts}
	\end{center}\vspace{-0.2cm}
\end{figure}

%% file: cvpr_sec_files_cr/boundedsupport.tex
\input{cvpr_sec_files_cr/fig_small_sup}

\vspace{-0.2cm}\subsection{Bounded support  matrices }\vspace{-0.1cm}\label{sec:compact-sup}
As stated above, much fewer layers are needed if we only attempt to fit transmission matrices sampled over a small area  of size  $\prng\times\prng$.


 When fitting layers, we need to account for the fact that  a point/plane illumination scattering through a volume is expanding. Thus, the aberration layers required to explain the transmission matrix are wider than the imaged area of size $\prng\times\prng$, see \figref{fig:support_hists}(a).
Expressing a transmission matrix with a compact support at the frequency basis is equivalent to illuminating the sample with plane waves spanning a wide set of angles, but all the waves pass through a narrow aperture of size $\prng\times \prng$. In this case most points inside the volume usually do not receive light from all angles.   In \figref{fig:support_hists}(c) we plot the narrow cone of angles arriving at three different points (a ray leaving a point $\bf{p_j}$ at angle $\dirvect$ is included in the cone only if this angle is not cropped by the aperture at the back of the sample). Since the cone of light reaching each point is narrower than the full numerical aperture, a local Fourier transform would not have content over the full butterfly area, but  cover an even lower range  of axial $\omega_z$ frequencies.  Hence, following the analysis in the previous section,  it can be sampled with fewer layers. We visualize the local spectra in the supplement. 

%% file: cvpr_sec_files_cr/fig_small_sup.tex
\begin{figure}[t!]
	\begin{center}
		\begin{tabular}{@{}c@{~~~}c@{~~~}c@{~~~}c@{}}	
			\includegraphics[height=0.19\textwidth]{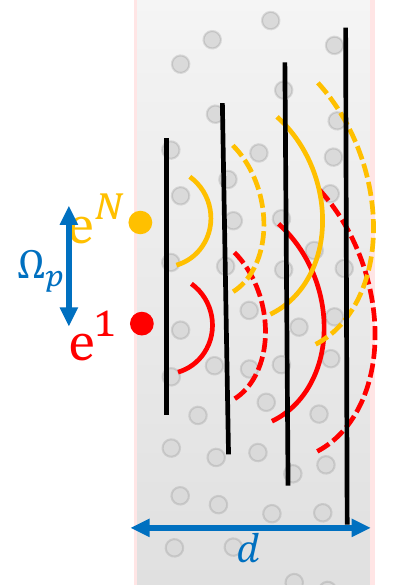}&
	        \includegraphics[height=0.19\textwidth]{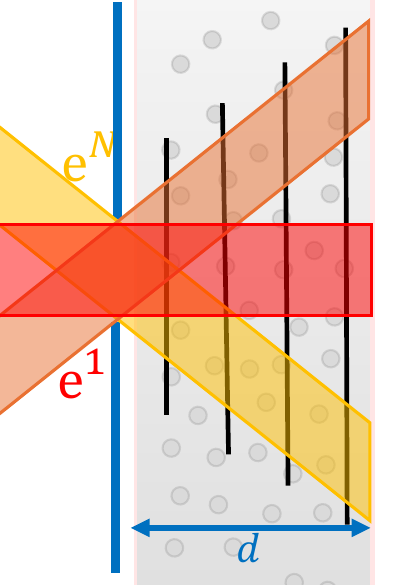}&
	         \includegraphics[height=0.19\textwidth]{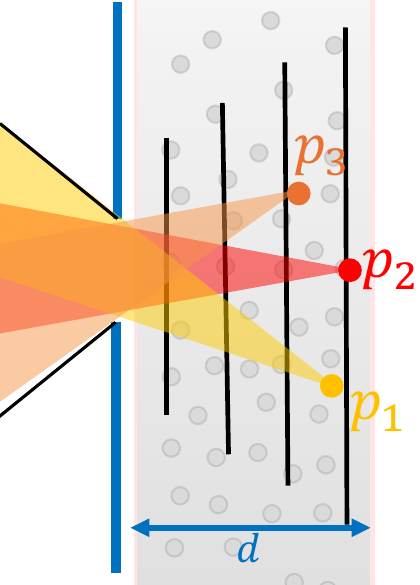}\vspace{-0.1cm}\\
	        	{\footnotesize{(a) Support primal}} &{\footnotesize{(b) Support freq.}} &{\footnotesize{(c) Local directions}} \vspace{-0.1cm}
	        	
		\end{tabular}
		\caption{ \footnotesize {\bf Layer support:} (a) As point sources expand while propagating through the volume, the width of the aberration layers should be wider than the measured support $\prng$. (b) To express compact support transmission matrices in the Fourier basis,  we illuminate the volume by a set of plane waves passing through an aperture.   (c) The local cone of illumination angles reaching different points inside the volume is much smaller than the actual range of incoming illuminations. 	}\label{fig:support_hists}\vspace{-0.2cm}
	\end{center}\vspace{-0.3cm}
\end{figure}

%% file: cvpr_sec_files_cr/discussion.tex

This paper studies how well we can fit transmission matrices with multi-slice models and   how many layers are  needed in practice.  We show that  the required number of layers is much lower than the prediction of Nyquist sampling theory. This is because the missing cone problem of 3D microscopy turns into an advantage when analyzing transmission matrices. Moreover, as tissue is forward scattering, its 3D spectrum contains most energy in the low frequencies and much less energy in the higher ones. Hence, even if the sparse sampling acts as a low-pass on the large axial frequencies, not much energy is lost.

We validate this prediction using both synthetic simulations and realistic lab measurements. We show that even though realistic transmission matrices contain backward and side-ward scattering which are not modeled by multi-slice matrices, they provide a reasonable approximation that is good enough for wavefront shaping focusing.



Our findings suggest that 
multi-slice models can be applicable for the design of non-local wavefront shaping systems, allowing for a volumetric correction with a small number of layered SLMs~\cite{Thaung:09,Wu:15,Laslandes:17,Furieri23}. Even if $2-3$ SLM layers may not be enough for correcting the full  field-of-view of a wide image through a thick tissue, they can allow us to increase the support of the fitted area by much more than $\times 2-3$ thus they can significantly accelerate a sequential wavefront-shaping correction of a wide image. 

\boldstart{Acknowladgments:} This research was funded by ERC SpeckleCorr-101043471, ISF 563/24.

\boldstart{Code and data:} can be obtained at \url{https://webee.technion.ac.il/people/anat.levin/papers/LayerdTmat\_cvpr25\_code.zip}

%% file: cvpr_sec_files_cr/missingcone_mc.tex

\section{Spectra coverage in the weakly scattering case  }
We start our derivation by considering a weakly scattering volume, expanding the analysis in Sec.~4  of the main paper. 
Our goal is to understand how  the reconstruction error is scaled as a function of the number of fitted layers.

As derived in the main paper under a weakly scattering  (first-Born) approximation the Fourier representation of the Transmission matrix corresponds to an entry from the Fourier transform of the RI volume. The wavefront scattering toward direction $\dirvect^v$ when illuminated by an incoming plane wave at direction $\dirvect^i$ is proportional, up to a multiplicative factor, to \cite{WOLF1969153,mertz2019introduction,Lauer2002}:
	\vspace{-0.1cm}{\small \BE\label{eq:single-scat-tm}
		\propto \int \refin(\ptd) e^{{2\pi i}{}{(\bunitkv^i-\bunitkv^v)\cdot \ptd}} d\ptd, \vspace{-0.1cm} \EE
	}with $\bunitkv^v=1/\lambda\dirvect^v$, $\bunitkv^i=1/\lambda\dirvect^i$.

As derived in the main paper, since only a subset of frequencies inside a butterfly shape can be measured, for an exact, error-free reconstruction, the minimal Fourier range and primal spacing are
\vspace{-0.2cm}{\small \BE\label{eq:opt-sampling}
	\repspacing^*=\NA\frng,\quad \eps^*=\frac{1}{\NA\frng}=\frac{\lambda}{\NA^2}.\vspace{-0.1cm}
	\EE } For a scattering volume of thickness $\smpthick$, the number of required layers is: 
\vspace{-0.1cm}{\small \BE\label{eq:Mopt-def}
	M^*=\frac{d \cdot \NA^2}{\lambda}.\vspace{-0.1cm}
	\EE}

In a physical wavefront shaping system we want to approximate the volume with a sparse set of slices, below this optimal bound. Our goal here is to analyze how much error is introduced by such sparse approximations.
For that, consider a scattering sample of thickness $\smpthick$, which we try to approximate using $M<M^*$ layers separated by $\eps=\frac{\smpthick}{M}$. We denote by $\errTap_M$ the reconstruction error of a given transmission matrix with the best $M$ layers
\vspace{-0.2cm}{\small \BE\label{eq:err-TM-M-layers}\errTap_M=\min_{\AbrVect_1,\ldots, \AbrVect_M}\|\TMcap-\TMml\|^2.\vspace{-0.1cm}\EE}

Our goal is to show that $\errTap_M$ decays relatively fast with $M$ and we can get a reasonable approximation to the transmission matrix even if the number of layers is significantly lower than the exact prediction in \equref{eq:Mopt-def}.
This is a result of two main properties: (i)~The volumes describing realistic tissue samples have more energy in the low frequencies; and (ii)~the structure of the missing cone implies that, in any case, a significant portion of the spectrum is not sampled by the transmission matrix.

To understand this, consider a naive selection of $M$ layers $\AbrVect_1,\ldots, \AbrVect_M$. Rather than actually solving an optimization problem,  we use the ground truth volume $\frefin(\bkv)$, and simply set to zero any frequency content of the RI volume  $\frefin(\bkv)$ at $\zkv$ values larger than the possible Nyquist range \vspace{-0.2cm}{\small\BE\label{eq:Nq-band-smp}\repspacing^M=\frac{1}{\eps}=\frac{\lambda M}{\smpthick},\vspace{-0.1cm}\EE}and we then Fourier transform  $\frefin(\bkv)$ to $\refin(\ptd)$ and sample planes at spacing $\eps=d/M$. The error of this approximation is basically the integral of content above the cut-off $\repspacing^M$. 
Since according to \equref{eq:Nq-band-smp}, the cut-off frequency  $\repspacing^M$ scales linearly with the number of layers, the portion of the spectrum which is lost by this low-pass operation scales linearly with the number of layers $M$. Thus, the naive answer is that the error of a multi-slice approximation decays linearly with $M$.
In practice we show below that the error decays much faster, since 
{\em large areas from the 3D spectrum of the sample are  not used by the transmission matrix. } 

To gain intuition,  consider \figref{fig:cone-parts}(b). 
We note that the butterfly shape is such that low 2D frequencies (i.e. low $\xykvnor$),  also span less content along the $\zkv$ axes, and hence these frequencies  are not lost even with low bandwidths $\repspacing^M$.
For the higher $\xykvnor$ frequencies, the lower $\zkv$ part, marked in dashed blue in \figref{fig:cone-parts}(b) is preserved, while the higher portion marked in red is lost.

Another important property of tissue is that its RI is locally smooth, and therefore if we look at its spectrum, we have much more energy in low frequencies than in high ones. Therefore, in the red areas of \figref{fig:cone-parts}(b) that are not sampled by the transmission matrix, there is less energy than in the lower frequencies. Thus, despite the fact that the multi-slice approximation sacrifices such frequencies, not much energy is being lost.

In most cases, it is hard to give analytic formulas for the decay of $\errTap_M$ as a function of $M$.
We can derive an analytic prediction in a simplified model where we assume that the spectrum of $\frefin(\bkv)$ has random values that are sampled from a uniform distribution for any frequency below a cutoff $\fsmprng$  (i.e. any frequency satisfying $\|\bkv\|\leq\fsmprng$), and zero content outside this band. Under this model we can prove that  $\errTap_M$ decays at least {\em quadratically} fast with $M$. This is a non-trivial result since the cut-off  frequency $\repspacing^M$ scales only {\em linearly} with $M$, see \equref{eq:Nq-band-smp}. Thus, if we just rely on Nyquist theory and compute the energy of the spectrum above the cut-off frequency, we  expect the error to decay linearly with $M$. The fact that the error decays quadratically with $M$  results from the butterfly structure.   However, we emphasize that this result uses the over-simplified assumption of a uniform content in $\frefin(\bkv)$. In the numerical simulation we see that this result is over-pessimistic, and with more realistic forward scattering volumes, the decay of $\errTap_M$ is significantly faster than a quadratic function. 

\vspace{-0.1cm}
\begin{claim}\label{claim:quad-decay}
	The reconstruction error of a transmission matrix corresponding to a  weakly scattering volume with a  uniform spectrum, is bounded by
	\vspace{-0.1cm}{\small \BE\label{eq:quadratic-err-curve}
		\errTap_M\leq \left(\frac{M^*-M}{M^*}\right)^2 \errTap_0,
		\vspace{-0.1cm}\EE	
	}where $\errTap_0$ is our ability to approximate a transmission matrix with no correction, namely with the ballistic light alone.\vspace{-0.1cm}
\end{claim}
We start with a numerical evaluation of this claim and then proceed to the proof.

\input{cvpr_sec_files_cr/fig_cone_parts_supp}
\input{cvpr_sec_files_cr/fig_numerical_sim_comp_spct_supp}

%
%

\boldstart{Numerical validation:}
To test the decay of the multi-slice fitting error as a function of the number of layers $M$, we generated  synthetic transmission matrices. We used two types of RI volumes, in the first case we selected random values for the spectrum $\frefin(\bkv)$, for any frequency $\|\bkv\|\leq\fsmprng$, using a band $\fsmprng=0.3/\lambda$. The second type of RI volume $\refin(\ptd)$ is filled with a set of spheres at random positions, and each sphere has a random RI different than that of the leading medium. This is a more realistic approximation to the structure of real tissue where we have cells with a certain RI embedded in a surrounding medium with a different RI. In \figref{fig:comp_spct}(a) we show a slice through the two spectra $\frefin(\bkv)$  we receive. With the random spheres the spectrum decays more naturally, and we have much more content at low frequencies than at high ones. This is a more realistic approximation to the spectrum of real tissue since the fact that tissue is forward scattering implies that its spectrum should have more content at lower frequencies.

We synthesized a target $\TMcap$ matrix using a multi-slice model where the planes are sampled very densely.  We generated a set of sparse multi-slice approximations with increasing $M$ values in two ways. First, we use gradient descent optimization to minimize the fitting error in \equref{eq:err-TM-M-layers}.  Second, we use a naive filtering of the ground truth volume, where we simply set to zero any frequency content at $\zkv$ values larger than the possible Nyquist range $\repspacing^M=\frac{ M}{\smpthick}$. We start with a low optical depth, and in \figref{fig:comp_spct}(b) we plot the square root of the reconstruction error as a function of $M$. One can see that with a uniform spectrum, 
the square root of the error  indeed decays linearly with $M$, suggesting that the actual reconstruction error decays quadratically with $M$.  The curves reach a plateau when $M$ exceeds the Nyquist requirements. This validates the prediction  of Claim \ref{claim:quad-decay}. 
However, with the more realistic forward scattering volume,  the decay of the fitting error is {\em significantly faster} than the analytic quadratic prediction.
For both types of volumes the optimization provides better fits with low $M$ values, but for high $M$ values it runs into local minima, and the fitting error it achieves can be higher than the one achieved by naive filtering. 

In our plots, the case $M=0$ refers to no correction at all, so effectively we assume that the transmission matrix is diagonal. The similarity between a diagonal transmission matrix and the target transmission matrix is a measure of the amount of ballistic light.

\input{cvpr_sec_files_cr/proof}

\section{Spectra coverage under multiple scattering }\label{sec:samplingtheory}

The  analytical analysis in the main paper  has assumed the sample is weekly scattering. 
 While it is hard to give analytical results in the case of multiple scattering, it appears that when the optical depth of the tissue is moderate, light paths undergo a small number of scattering events without being completely scrambled, and the same results hold. The reason is that, since tissue is forward-scattering, light only scatters at small angles. Therefore, after a small number of scattering events, most light paths remain within the butterfly area of the spectrum.
This is usually the regime that wavefront-shaping algorithms like~\cite{Dror22,DrorNatureComm24} attempt to tackle as such algorithms attempt to push the depth at which conventional microscopes can see, but they do not yet attempt to image extremely deep where scattering is fully diffused.
\input{cvpr_sec_files_cr/fig_stats_3D_paths}

To illustrate this, we carry a Monte-Carlo path tracing through a volume. 
We consider light paths of the form $\vec{\ptd}=\ptd_0,\ptd_1,\ldots,\ptd_\ell$, and denote the direction of the ray between points $\ptd_k,\ptd_{k-1}$ by $\dirvect_k$. 
That is, when a light path scatters at a point $\ptd_k$ it is changing its direction from $\dirvect_k$ to $\dirvect_{k+1}$, see illustration in \figref{fig:path_stats}(a).
This is equivalent to sampling the 3D frequency $\bkv=1/\lambda(\dirvect_{k+1}-\dirvect_{k})$. 
Our  Monte-Carlo simulation samples the first direction on the path anywhere inside the numerical aperture of the expected illumination objective. We  sample paths according to a target phase function and optical density but discard all paths that eventually exit the volume at directions outside the numerical aperture as such paths cannot be collected by the imaging objective. However, the inner nodes on the path can include scattering at large angles beyond the $\NA$.
After tracing multiple paths through the volume we compute a histogram of the traced 3D frequencies and we visualize a $\xkv,\zkv$ slice through it. 
In \figref{fig:path_stats}(b-d) we plot a few such histograms. We start by setting the optical depth of the volume to $\od=0.5$ (\figref{fig:path_stats}(b)), in this case most paths scatter only once and indeed most of the histogram content is inside  the butterfly shape 
of \figref{fig:cone-parts-proof}(a).
There is some content outside this ellipse since a few of the paths we traced are longer. In \figref{fig:path_stats}(c) we repeated a similar simulation increasing to $\od=5$, meaning that the average number of scattering events on a path is $5$. Such longer paths turn in a variety of directions, and we get content over the entire 3D spectrum. In \figref{fig:path_stats}(b,c) we have assumed that the light is scattering isotropically (Henyey-Greenstein parameter $g=0$). In \figref{fig:path_stats}(d) we repeated the experiments with $\od=5$ but used a forward scattering phase function which better describes tissue, with a Henyey-Greenstein parameter $g=0.9$. Unsurprisingly, most of the paths traced contribute to {\em low frequencies}, and even though  paths are long, most content is inside the butterfly and not in the area of the missing cone. This observation has a significant impact on the analysis of layered approximations: even if they can only capture the low frequencies of the refractive volume, they can still provide a good approximation to the transmission matrix.  

\subsection{Path tracing with compact support}

Due to the large memory requirements of 4D transmission matrices, they are usually only sampled inside a bounded range. That is, we only move illumination point sources inside a small area  of size  $\prng\times\prng$, and measure a bounded set of columns.

When fitting layers, we need to account for the fact that  a point/plane illumination scattering through a volume is expanding. Thus, the aberration layers required to explain the transmission matrix are wider than the imaged area of size $\prng\times\prng$, see \figref{fig:support_hists}(a).
Such compact support transmission matrices can  be fitted with a sparser set of layers. To see this we consider \figref{fig:support_hists}(b). Expressing a transmission matrix with a compact support at the frequency basis is equivalent to illuminating the sample with plane waves spanning a wide set of angles, but all the waves pass through a narrow aperture of size $\prng\times \prng$. In this case most points inside the volume usually do not receive light from all angles.   In \figref{fig:support_hists}(c) we plot the narrow cone of angles arriving at three different points (a ray leaving a point $\bf{p_j}$ at direction $\dirvect$ is included in the cone only if this ray is not cropped by the aperture at the back of the sample). Since the cone of light reaching each point is narrower than the full numerical aperture, a local Fourier transform would not have content over the full butterfly area, but  cover an even lower range  of axial $\omega_z$ frequencies.  Hence, following the analysis in the previous section,  it can be sampled with fewer layers. 

To illustrate this, we use again Monte-Carlo path tracing. In \figref{fig:support_hists}(d) we plot the histogram of frequencies visited by a M.C. process, but we only record the paths that passed in three local regions marked in \figref{fig:support_hists}(c). Comparing these histograms to the histogram of paths in the entire volume, we see that the local histograms have a much narrower axial range.

As another way to understand it, we show in Fig. 4 of the main paper
an xz slice from the ground truth RI volume   and from a few reconstructions. We first optimized for a layer fitting with a dense  sampling (high layer number $M$). Even in this case, the axial resolution of the reconstruction is poor, and  {\em the axial resolution reduces} when the support  $\prng$ is low. This explains why lower supports can be fitted with fewer layers. For each of the two supports we also show a sparse fit,
with the minimal  $M$ value that provides a good prediction of the transmission matrix.

\input{cvpr_sec_files_cr/fig_small_sup_with_spect}

%% file: cvpr_sec_files_cr/fig_cone_parts_supp.tex
\begin{figure}[t!]
	\begin{center}
		\begin{tabular}{@{}c@{~~~~~}c@{}}	
			\includegraphics[height=0.2\textwidth]{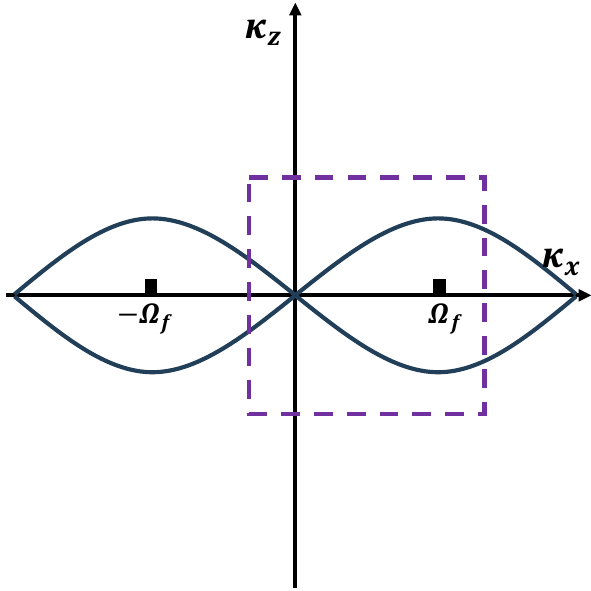}&
			\includegraphics[height=0.2\textwidth]{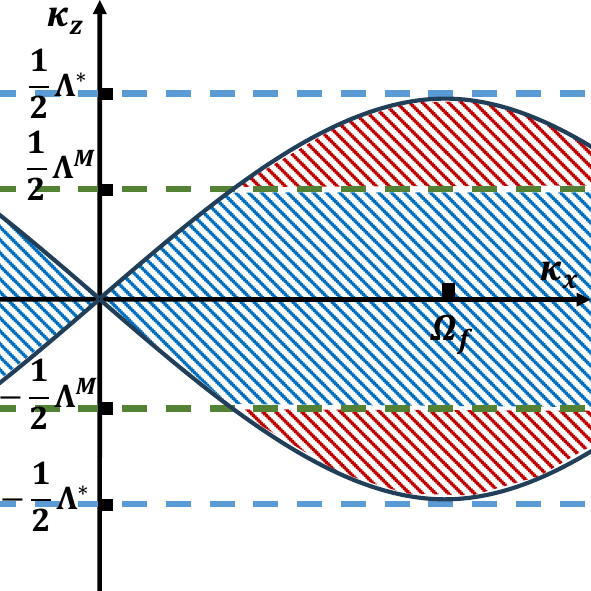}
			\\(a)&(b)
		\end{tabular}\vspace{-0.2cm}
		\caption{\footnotesize {\bf Spectrum structure:} (a) An $xz$ slice out of the spectrum of   $\frefin(\bomg)$. Entries of a weakly-scattering  transmission matrix limited by an aperture $\NA$ only lie inside the butterfly area. (b) Zooming on the center right area of (a) (purple square). Assuming the $\omega_z$ axis is cut at $\pm \frac12 \repspacing^M$, the transmission matrix entries inside the dashed blue area are maintained, and the entries in the dashed red area are lost. 	}\label{fig:cone-parts}
	\end{center}\vspace{-0.2cm}
\end{figure}

%% file: cvpr_sec_files_cr/fig_numerical_sim_comp_spct_supp.tex
\begin{figure*}[t!]
	\begin{center}
		\begin{tabular}{@{}c@{~~}c@{~}c@{}}	
			\includegraphics[height=0.17\textwidth]{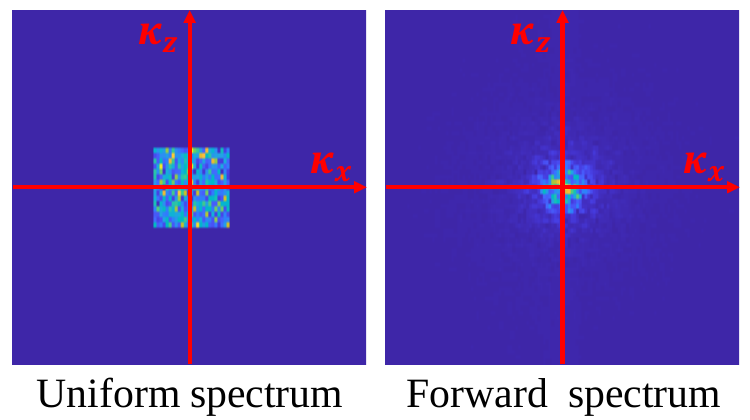}&
			\includegraphics[height=0.17\textwidth]{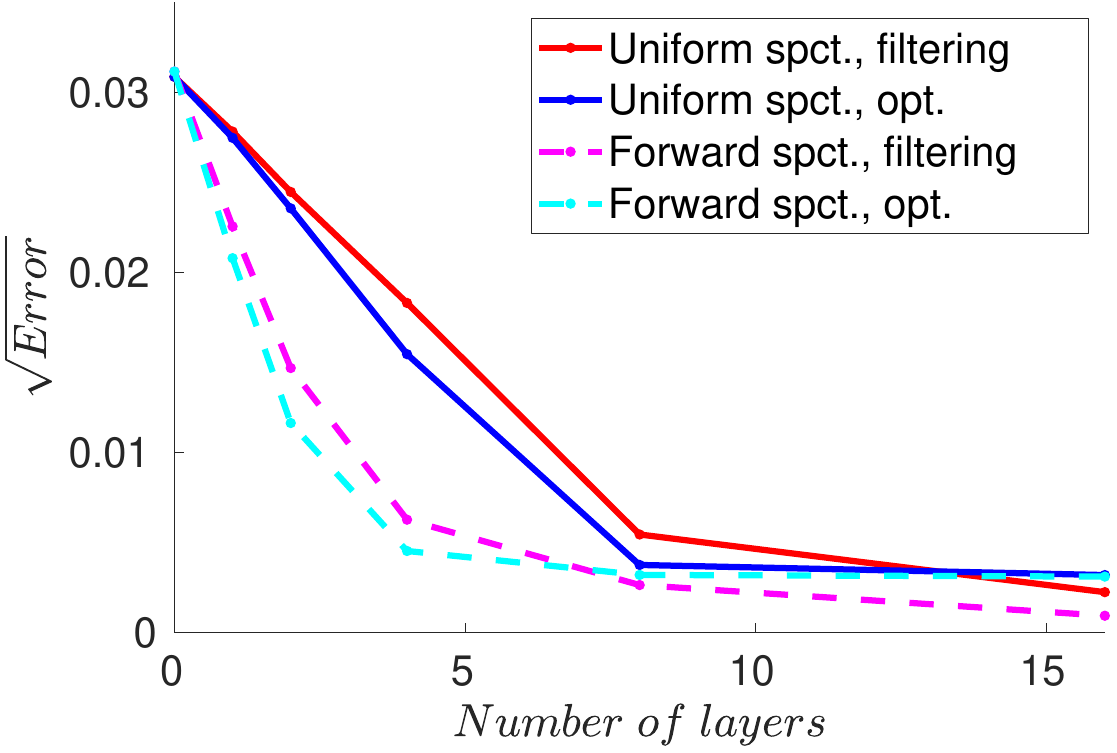}&
			\includegraphics[height=0.17\textwidth]{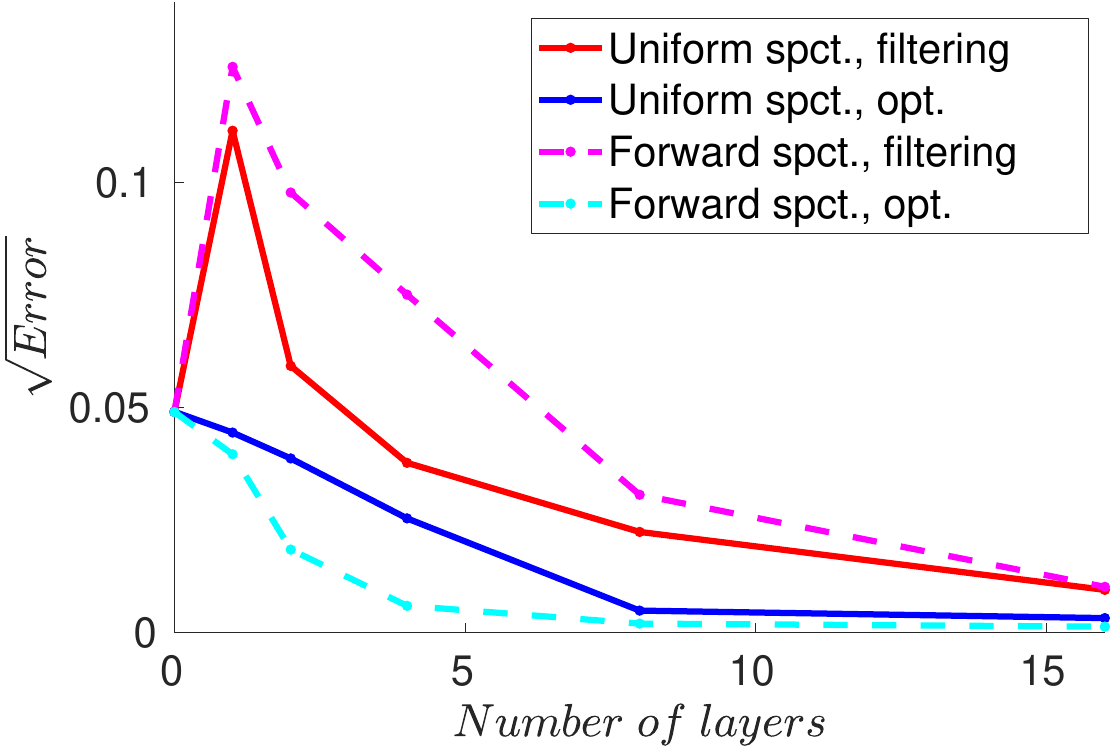}\\
{\footnotesize	(a) Spectra} &{\footnotesize(b) $\od=0.5$}&	{\footnotesize 	 (c) $\od=4$}
		\end{tabular}\vspace{-0.2cm}

		\caption{\footnotesize {\bf Numerical simulation:} We synthesize transmission matrices using a very dense multi-layer model, and test how well we can fit them with a sparser set of layers. We compare two types of volumes, whose spectra are illustrated in (a). The first one has a random spectrum sampled from a uniform distribution, and the second  corresponds to a forward scattering spectrum with more content at the low frequencies.
			 We plot the square root of the fitting error as a function of the number of layers. As predicted in claim \ref{claim:quad-decay}, with a uniform spectrum this decays linearly with the number of layers until we pass the Nyquist limit. The error with the more realistic forward scattering spectrum decays much faster, but analytic characterization is harder.
			 We  compute the layers using a naive filtering of the ground truth as well as using gradient descent optimization. The naive filtering provides good results at low optical depths, as illustrated in (b), but fails at higher $\od$s as illustrated in (c).	}\label{fig:comp_spct}

	\end{center}\vspace{-0.2cm}
\end{figure*}

%% file: cvpr_sec_files_cr/proof.tex
\subsection{Deriving the multi-slice fitting error}
Below we proof Claim 1. 
%

\boldstart{Proof:}
To prove this result we use a brute-force selection of the layers $\AbrVect_1,\ldots, \AbrVect_M$.  We simply set to zero any frequency content of the refractive volume  $\frefin(\bkv)$ at $|\zkv|$ values larger than the possible Nyquist range \BE\label{eq:Nq-band-smp-2}\repspacing^M=\frac{1}{\eps}=\frac{\lambda M}{\smpthick},\EE and we then Fourier transform  $\frefin(\bkv)$ to $\refin$ and sample planes at spacing $\eps$.
In the following paragraphs we offer an upper-bound calculation of the number of transmission matrix entries that lie outside the band $\repspacing^M$  and are hence lost by the low-pass operation. We show this scales quadratically with $M$.

To perform this calculation we consider \figref{fig:cone-parts-proof}. 
We note that the butterfly shape is such that low 2D frequencies (i.e. low $\xykvnor$ ) also span less content in the $\zkv$ axes, and hence these frequencies  are not lost even with low bandwidths $\repspacing$.
For the higher $\xykvnor$ frequencies, the lower $\zkv$ part marked in blue is preserved, while the higher portion marked in red  in \figref{fig:cone-parts-proof}(b) \comment{\figref{fig:cone-parts}(b)} is lost. Below we preform a conservative calculation of how many transmission matrix entries are included in the red area, and show that this number is bounded by a quadratic function of $(M^*-M)$.
\input{cvpr_sec_files_cr/fig_cone_parts_proof_supp}
Using the assumption that the frequency content of the volume is smaller than the cut-off frequency set by the numerical aperture ($\fsmprng<\frng$) we can approximate the butterfly boundaries at areas where the volume has content as linear curves:
\BE\label{eq:butterfly-shape-approx}
|\zkv|\leq \NA \xykvnor.
\EE
Also, if $\fsmprng<\frng$ the maximal $\zkv$ frequency at which we observe content does not get to $\NA\frng/2$ as derived in Eq. 8 of the main paper, but can be actually reduced to $\NA\fsmprng$.
Thus, for error-free reconstruction it is enough to maintain content up to a cut-off frequency of \BE\frac12\repspacing^*=\NA\fsmprng,\EE and as a result the minimal number of required layers is
\BE\label{eq:M-opt}
M^*={2\NA \fsmprng \smpthick}
\EE
With this model we treat the red area as a triangle, and we denote the frequency at which the triangle intersects with the z-axis bandwidth as $\kscl^M$, it can be shown that
\BE \label{eq:cutoff-M} \kscl^M=\frac{\repspacing^M}{2\lambda\NA}=\frac{ M}{2\NA\smpthick}.\EE 
To compute the number of transmission matrix entries in the red region, we need to compute an integral of the following form
\BE
\int_{\kscl=\kscl^M}^{\fsmprng} \NA(\kscl-\kscl^M)\cdot(2\pi\kscl)\cdot f(\kscl) d\kscl.
\EE
The first term in this integral is the distance of the triangle from the cutoff frequency (the triangle at frequency $\xykvnor=\kscl$ spans $\zkv$ values in a range from $\NA\kscl^M$ to $\NA\kscl$ ), the 2nd term encodes the fact that in the 3D Fourier domain there is a full circle of frequencies with norm $\xykvnor=\kscl$. Finally the function $f(\bkv)$ encodes the density  of transmission matrix entries which are mapped to frequency  $\bkv=(\xkv,\ykv,\zkv)$. Below we show this is bounded by
\BE
f(\bkv)\leq\frac{c}{\xykvnor}
\EE
where $c$ is a constant scalar.
With this we can bound the volume of missing frequencies as
\BE
\int_{\kscl=\kscl^M}^{\fsmprng} \NA(\kscl-\kscl^M)\cdot(2\pi)\cdot c d\kscl=\NA\pi c \left(\fsmprng-\kscl^M\right)^2
\EE
By plugging $\fsmprng$ from \equref{eq:M-opt}, and the value for $\kscl^M$ as in \equref{eq:cutoff-M} we get that the density of transmission matrix entries in the filtered red volume is proportional to 
\BE
\int_{\omega=\omega^M}^{\frac12\fsmprng} \NA(\omega-\omega^M)\cdot(2\pi)\cdot c d\omega \propto (M^*-M)^2.
\EE
Using the same reasoning when we have $M=0$ layers the error in reconstructing the transmission matrix, namely the full volume of the butterfly shape is proportional $(M^*)^2$. This leads us to the desired \equref{eq:quadratic-err-curve}.

The last thing we need to prove is that the density of transmission matrix entries around the 3D frequency $\bkv=(\xkv,\ykv,\zkv)$ is bounded by
\BE
f(\bkv)\leq\frac{c}{\xykvnor}
\EE
For that we recall that we sample a frequency $\bkv$ when we have illumination and viewing directions whose frequencies satisfy $\bunitkv^i-\bunitkv^v=\bkv$, where $\bunitkv^i=1/\lambda\dirvect^i,\bunitkv^v=1/\lambda\dirvect^v$, so $\bunitkv^i,\bunitkv^v$ are vectors of norm $1/\lambda$.
For that consider  illumination and viewing directions which we parameterize using a 2nd order approximation as
\BEA
\!\!\bunitkv^i&=&\!\left(\!\!\begin{array}{c}\tau_x+\xkv\\\tau_y+\ykv\\\frac1\lambda-\frac\lambda2((\tau_x+\xkv)^2+(\tau_y+\ykv)^2)\end{array}\!\!\right) \\ \bomg^v\!&=&\!\left(\!\!\begin{array}{c}\tau_x\\\tau_y\\\frac1\lambda-\frac\lambda2(\tau_x^2+\tau_y^2)\end{array}\!\!\right)
\EEA 
The difference between these two vectors in the first two coordinates is $\xkv,\ykv$. The difference in the 3rd coordinate can be expressed as 
\BE
\begin{split}
\zkv &=\lambda(\tau_x\xkv+\tau_y\ykv)-\frac\lambda2 \xykvnor^2 \\
&= \lambda((\tau_x,\tau_y)\cdot (s^1\xykvnor))  -\frac\lambda2 \xykvnor^2
\end{split}
\EE
where in the right hand side of the above equation we use the 2D unit norm vector $s^1=(\xkv,\ykv)/ \xykvnor$. 
With this notation we see that to get the 3D frequency $\zkv$ we need to use viewing directions whose $\tau_x,\tau_y$ satisfy the linear constraint
\BE ((\tau_x,\tau_y)\cdot s^1)=\frac{1}{\xykvnor}\left(\zkv+ \frac12 \xykvnor^2\right).\EE
The density of directions satisfying this constraint scales as $1/\xykvnor$.

\eop

%% file: cvpr_sec_files_cr/fig_cone_parts_proof_supp.tex
\begin{figure*}[t!]
	\begin{center}
		\begin{tabular}{@{}c@{~~~~~}c@{~~~~~}c@{}}	
			\includegraphics[height=0.25\textwidth]{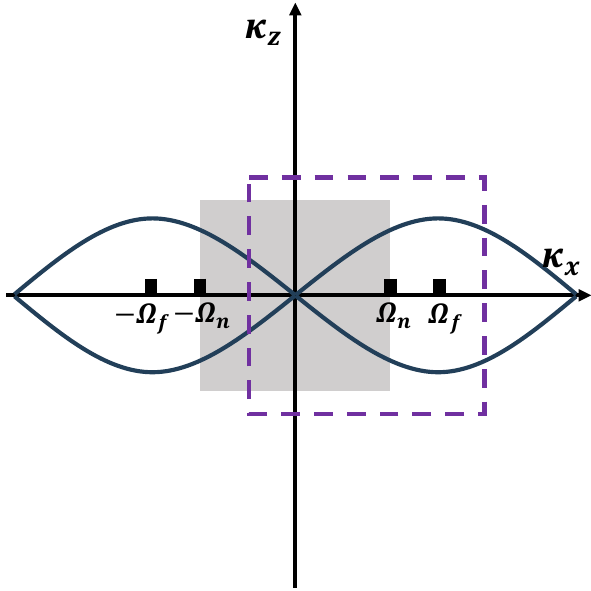}&
			\includegraphics[height=0.25\textwidth]{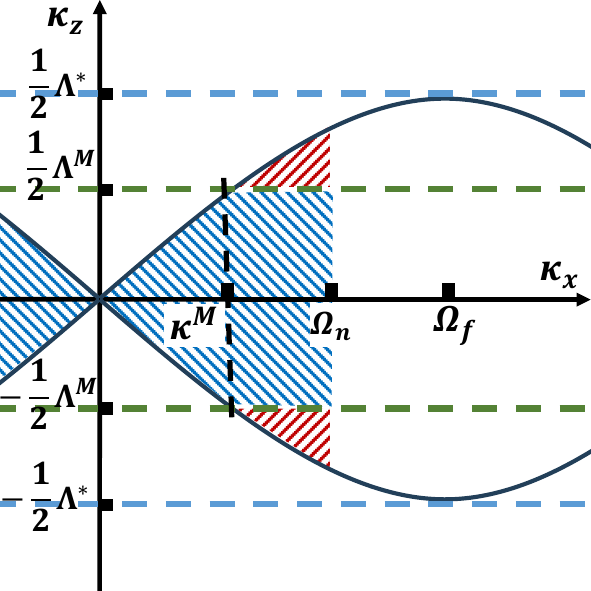}&
			\includegraphics[height=0.25\textwidth]{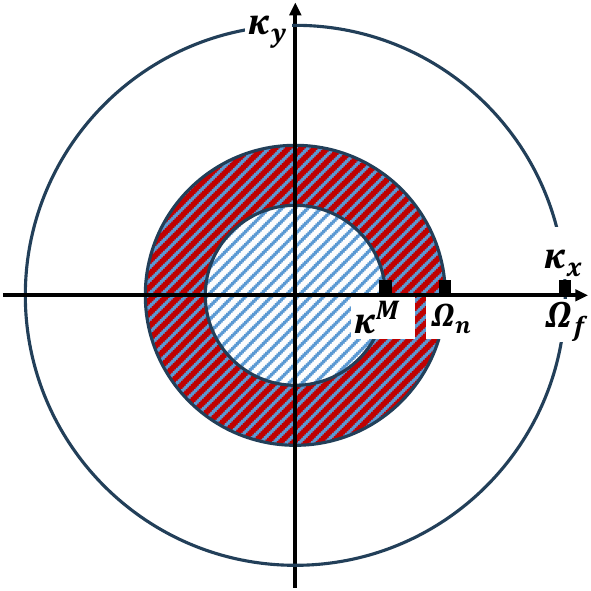}
			\\
			(a) & (b) &(c)
		\end{tabular}
		\caption{\footnotesize {\bf Spectrum structure:} (a) an $x-z$ slice out of the spectrum of   $\frefin(\bomg)$. Entries of the transmission matrix limited by an aperture $\NA$ only lie inside the butterfly area. We also assume the content of  $\frefin(\bomg)$ is limited to a bend of support $\fsmprng$ marked in gray in the figure. (b) zooming on the center right area of (a) (purple rectangle), assuming the $\omega_z$ axis is cut at $\pm \frac12 \repspacing$, the transmission matrix entries inside the dashed blue area are maintained, and the entries in the dashed red area are lost. The approximation error with $M$ layers is proportional to the red volume, which is shown to scale quadratically with $M$. (c) The $x-y$ projection of the maintained/lost areas.  	}\label{fig:cone-parts-proof}
	\end{center}
\end{figure*}

%% file: cvpr_sec_files_cr/fig_stats_3D_paths.tex
\begin{figure*}[t!]
	\begin{center}
		\begin{tabular}{@{}c@{}}	
			\includegraphics[height=0.3\textwidth]{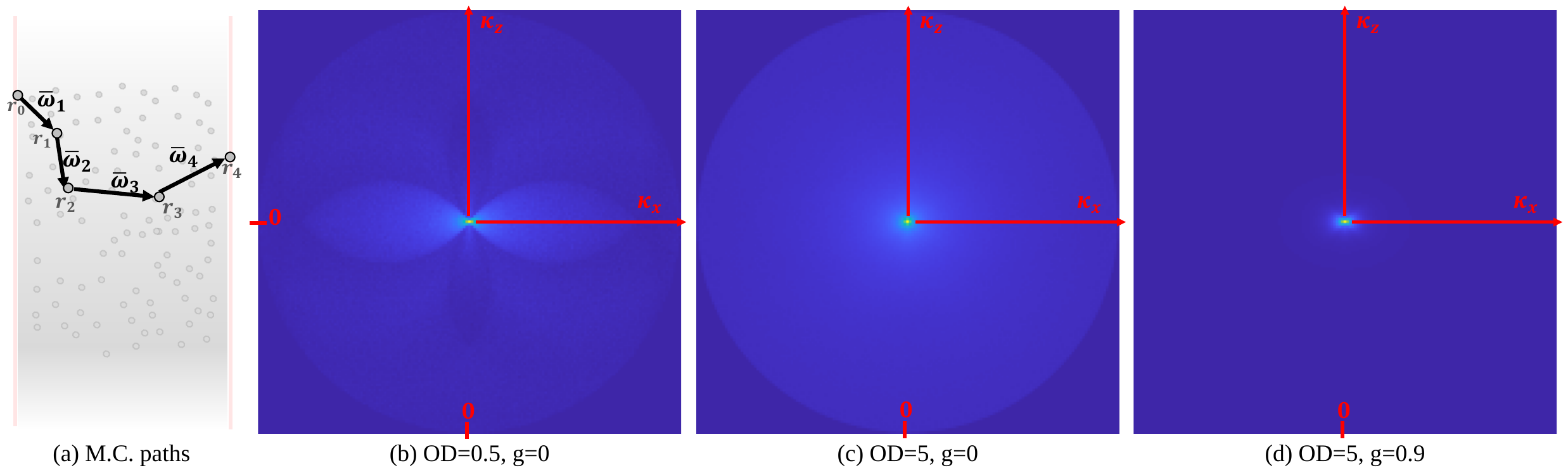}
		\end{tabular}
		\caption{\footnotesize {\bf 3D frequencies of MC paths:} (a) An illustration of a Monte-Carlo path in the volume. (b-d) We plot a histogram of the 3D frequencies traced by such paths.  (b) A weakly scattering volume of $\od=0.5$ with an isotropic phase function, shows most frequencies lie inside a narrow butterfly area.  (c) When $\od$ is increased to $5$ multiple scattering paths generate content at other frequencies. (d) Even at high optical depths, if the phase function is forward scattering as in tissue, most content is at the low frequencies.  }\label{fig:path_stats}
	\end{center}
\end{figure*}

%% file: cvpr_sec_files_cr/fig_small_sup_with_spect.tex
\begin{figure*}[t!]
	\begin{center}
		\begin{tabular}{@{}c@{~~~}c@{~~~}c@{~~~}c@{}}	
			\includegraphics[height=0.19\textwidth]{figs/layer_support/layeres_support1.pdf}&
	        \includegraphics[height=0.19\textwidth]{figs/layer_support/layeres_support2.pdf}&
		    \includegraphics[height=0.19\textwidth]{figs/layer_support/layeres_support3.pdf}&
			\includegraphics[height=0.19\textwidth]{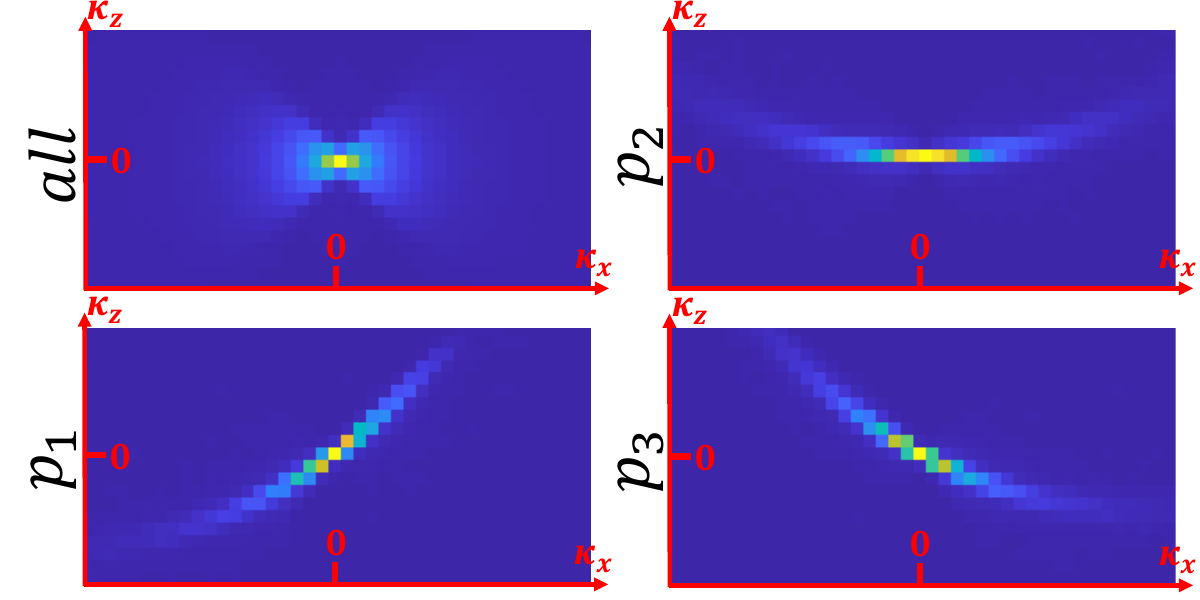}
		\\
	{\small{(a) Support primal}} &{\small{(b) Support freq.}} &{\small{(c) Local directions}} &{\small{(d) Path histograms}}
		\end{tabular}
		\caption{ \footnotesize {\bf Layer support:} (a) as point source expand while propagating through the volume the width of the aberration layers should be wider than the support $\prng$ over which the transmission matrix is measured. (b) To express compact support transmission matrices in the Fourier basis,  we illuminate the volume by a set of plane waves passing through an aperture. Such waves also expand through the volume to an area wider then the aperture.  (c) The local cone of illumination angles reaching different points inside the volume is much smaller than the actual range of incoming illuminations. (d) Since locally each point receives light through a limited angular cone, the local Fourier transform has a lower axial range. To show this we plot the full histogram of angles scanned by Monte-Carlo paths (this is a zoom of the histograms in \figref{fig:path_stats}). We also plot only the histogram of paths passing through the 3 points marked in (c). One can see that such local path-histograms have a limited axial spread.   Due to this limited range they can be explained with fewer layers. 	}\label{fig:support_hists}
	\end{center}
\end{figure*}

%% file: cvpr_sec_files_cr/empirical_fits_supp.tex
\input{cvpr_sec_files_cr/fig_fit_allMats_supp}
\input{cvpr_sec_files_cr/fig_fit_chicken_varysup_supp}

\section{Empirical evaluation of  multi-slice approximations}
In this section we consider a few transmission matrices and check empirically how well we can approximate them with a multi-slice model.
We start with numerically simulated transmission matrices using a more accurate wave-propagation model. 
We then test lab-captured transmission matrices measuring realistic scattering samples, including thick multiple scattering examples.  

While the analysis used the Fourier representation of the transmission matrix,  our simulations and measurements use transmission matrices expressed in the primal domain, since in practice, wavefront shaping algorithms use primal measurements.

%
%
%

\subsection{Monte-Carlo transmission matrices}
The transmission matrices used in the simulations of the previous sections were generated using multi-slice models with very dense slices. The multi-slice model is only an approximation to the full wave-equation because it does not simulate back-scattering paths. To test the discrepancy between this model and the full wave equation,   
we used the Monte-Carlo algorithm of \cite{Bar:2019,Bar:2020} to synthesize transmission matrices. It has been shown that this algorithm generates complex fields with physically correct statistics which are equivalent to an exact solution to the wave equation, yet it is much faster to compute and scales to much larger scenes. In particular, the M.C. algorithm simulates scattering from particles at any point in the volume (not only on sparse slices) and at all angles (including back-scattering). 
We simulated a volume of thickness $\smpthick=50\mu m$  at $\lambda=0.5\mu m$ and illuminated it with point sources spaced over an area of $\prng=18\mu m$. We measured the scattered fields over a wider support of $30\times 30 \mu m$. We used a forward scattering phase function as is common in real tissue. We simulated volumes  with two different optical depths and fitted them with a multi-slice model, while increasing the number of layers. The fit results are plotted in \figref{fig:fitting-all}(top). The layered transmission matrices have $\times 7$ lower fitting error when compared to the ballistic term alone which is equivalent to fitting with zero layers. However, even when the number of layers increases and approaches the Nyquist limit, the fitting error does not decrease to zero. 
This  failure is a combination of two issues. First, the fact that the actual transmission matrix includes scattering at wide angles which are not modeled by the layered approximation, and second, the fact that the optimization problem is not convex, and the gradient descent does not converge to a global optimum. 

As another way to understand the quality of the fit, we tested the correlation between columns of the exact and fitted transmission matrices and measured
\BE\label{eq:cor-TM-M-layers}\corrTap_M=\frac{1}{K}\sum _k \left|\frac{ {\bt^k_{\text{exact}}}^T  \cdot \bt^k_{\text{fit,M}}}{\|\bt^k_{\text{exact}}\|\cdot \|\bt^k_{\text{fit,M}}\|}\right|^2
\EE
where $\bt^k_{\text{exact}},\bt^k_{\text{fit,M}}$ are columns from the input and fitted transmission matrices. 
This provides a prediction of the percentage of energy we can deliver to a point behind the tissue if we use $\bt^k_{\text{fit,M}}$ as a wavefront shaping correction, rather than the exact  $\bt^k_{\text{exact}}$ (note that this only evaluates the energy with respect to the modes included in the input transmission matrix, but there may be additional correction modes not captured by the transmission matrix. Namely, if we would measure a transmission matrix over a wider $\prng$ support we could focus more light to a point).
With sufficient $M$ values we can deliver more than $80\%$ of the energy, as plotted in the top row of \figref{fig:fitting-all}(b). 
In \figref{fig:fitting-all}(c) we also show some examples of the spot behind the tissue if we use 
$\bt^k_{\text{fit,M}}$ as a wavefront shaping correction. For $M\geq 5$ layers the fit is good enough to provide a sharp spot.
Note that while we only show focusing at one point, the layers are optimized such that they allow us to focus at {\em any point} inside the $\prng\times \prng=18\times 18\mu m$ window.

\subsection{Acquired transmission matrices}
We used the Hadamard algorithm of~\cite{PopoffPhysRevLett2010} to capture transmission matrices of real samples in the lab. We measured a layer of parafilm, and two  slices of mouse brain and  chicken-breast tissue. We measured the parafilm layer  to be of thickness $\smpthick=46\mu m$, the mice brain to have thickness $400\mu m$ and the chicken breast to be of thickness $\smpthick=170\mu m$. The transmission matrices cover an area of $\prng=25\mu m$. The measurement is very noisy, mostly due to vibrations during the long capture.  Due to the noisy acquisition 
the fit is not as good as in the synthetic case, but the fitted wavefronts  can still focus more than $50\%$ of the energy and generate sharp spots behind the tissue.
We show the fits for the parafilm and brain samples  in the two lower panels of \figref{fig:fitting-all}. 
We also show the spot we can get behind the tissue using the approximated transmission matrix. Note that these wavefronts are computed numerically, by multiplying the approximated wavefront by the captured transmission matrix.


In \figref{fig:fitting-chicken-varysup} we show fits on the chicken breast matrix. Here we tested the quality of the fit as a function of ranges $\prng$. For that, the algorithm attempts to fit a subset of the measured columns, limited into smaller spatial ranges $\prng$. As in Fig. 6 of the main paper, we see that smaller spatial ranges can  be fitted with fewer layers, since if the transmission matrix only covers a limited spatial support, the range of illumination angles reaching each point in the volume is effectively very narrow, hence locally, the Fourier transform has a limited axial range.
%
From the results in \figref{fig:fitting-chicken-varysup} we can see that for the $25\times 25\mu m$ support we measured we can obtain good focusing with about $3-4$ layers. 
	Fitting a full field-of-view of several hundred microns likely requires additional layers. However, we observe that while using $M=1$ allows us to focus on a small area of about $2\times 2\mu m$, with $M=2$ focus is expanded  to a $\times 9$ larger area of $6\times 6\mu m$, and with $M=3$, we can focus on an area $\times 150$ larger, reaching  $25\times 25\mu m$. Therefore, using multiple layers can significantly accelerates a sequential scanning of a wide field-of-view.
 
%

Mice brain samples used in this manuscript were approved by Institutional Animal Care and Use Committee (IACUC) at the Hebrew University of Jerusalem (MD-20-16065-4).

%% file: cvpr_sec_files_cr/fig_fit_allMats_supp.tex
\begin{figure*}[t!]
	\begin{center}
		\begin{tabular}{@{}c@{~~~~~~~}c@{}}	
			\begin{tabular}{@{}c@{~~~~}c@{}c@{}}	
			{\raisebox{1.90cm}{\rotatebox[origin=c]{90}{Monte Carlo }}}&
			\includegraphics[height=0.21\textwidth]{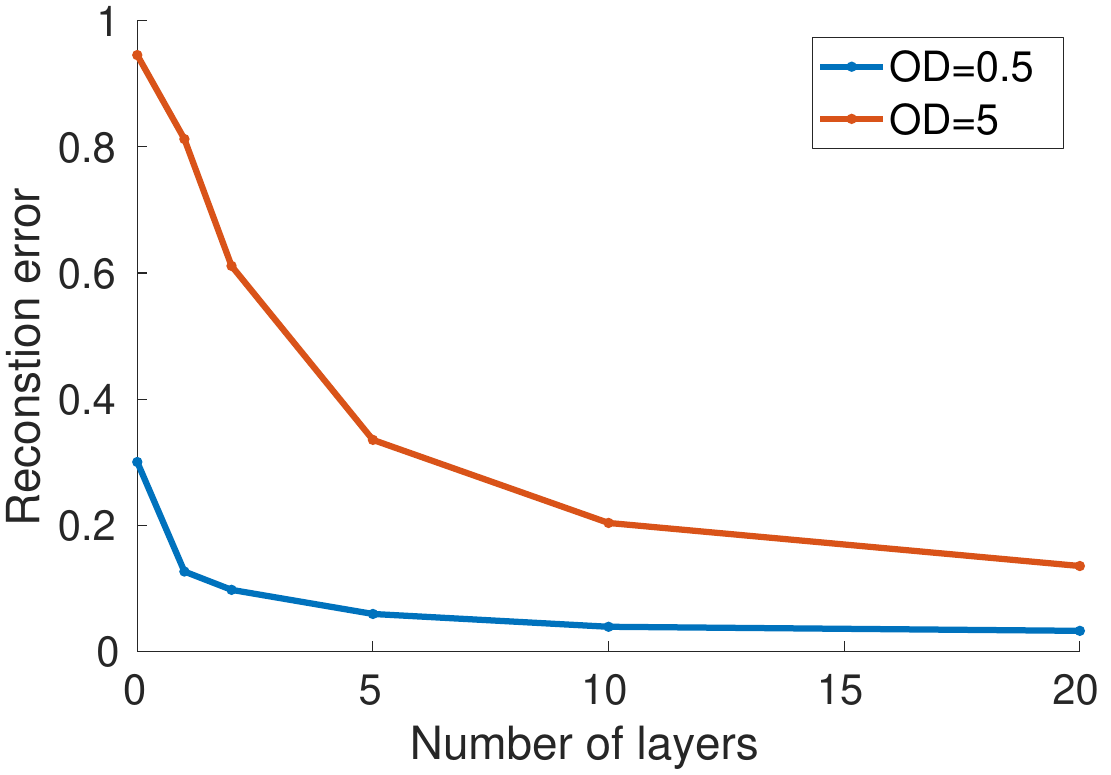}&
			\includegraphics[height=0.21\textwidth]{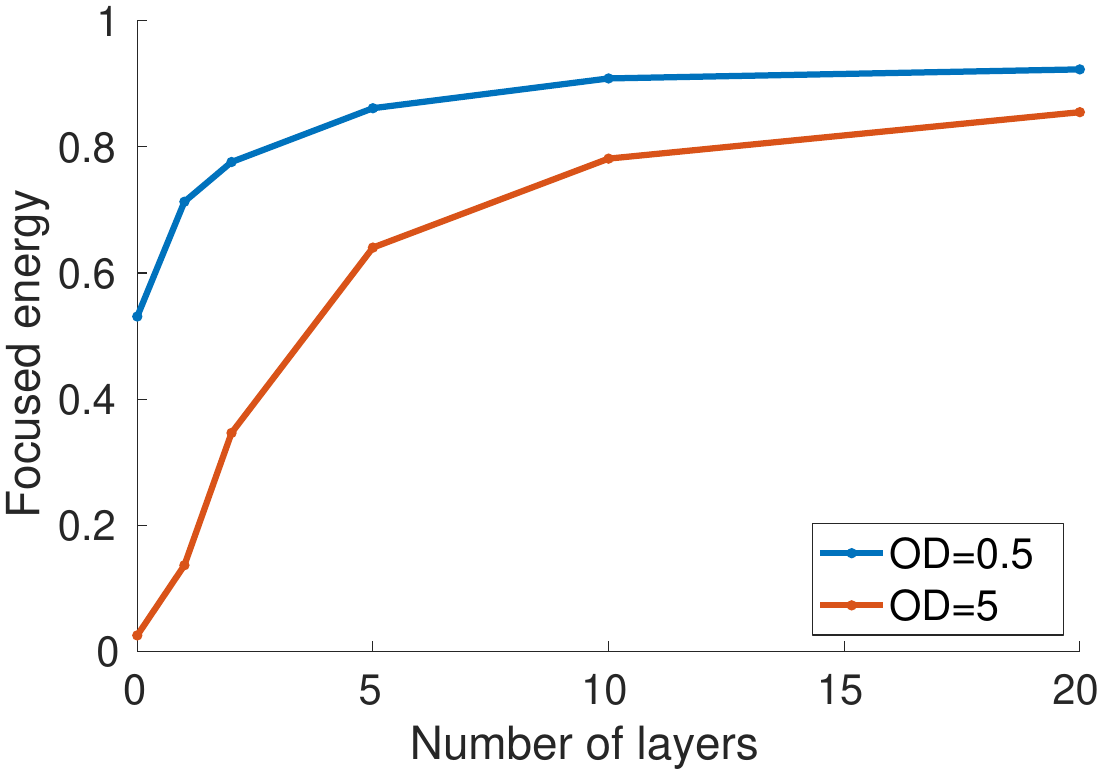}
				\end{tabular}&
				\begin{tabular}{@{}c@{}c@{}c@{}@{}c@{}c@{}c@{}}	
				\includegraphics[height=0.08\textwidth]{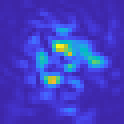}&
				\includegraphics[height=0.08\textwidth]{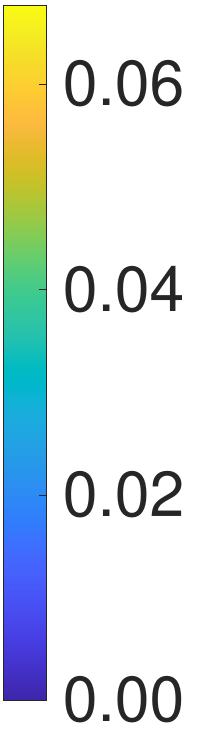}&
				\includegraphics[height=0.08\textwidth]{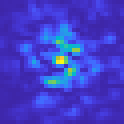}&
				\includegraphics[height=0.08\textwidth]{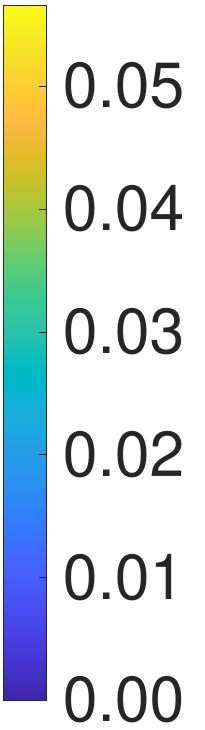}&
				\includegraphics[height=0.08\textwidth]{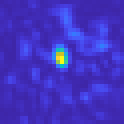}&
				\includegraphics[height=0.08\textwidth]{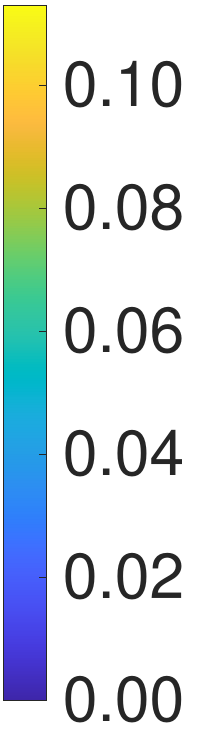}\vspace{-0.2cm}\\
				$M=0$&&$M=1$&&$M=2$&\\
				\includegraphics[height=0.08\textwidth]{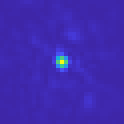}&
				\includegraphics[height=0.08\textwidth]{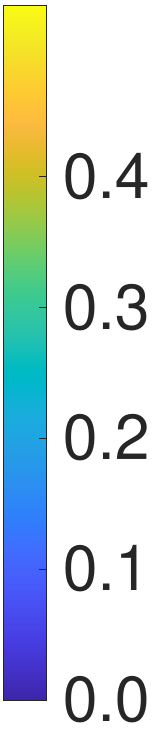}&
				\includegraphics[height=0.08\textwidth]{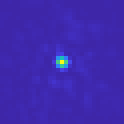}&
				\includegraphics[height=0.08\textwidth]{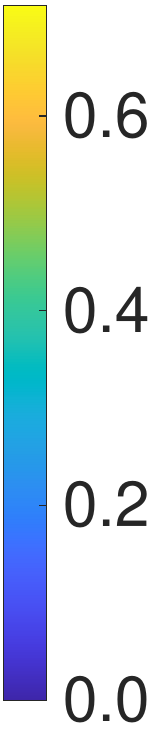}&
				\includegraphics[height=0.08\textwidth]{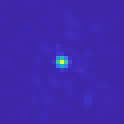}&
				\includegraphics[height=0.08\textwidth]{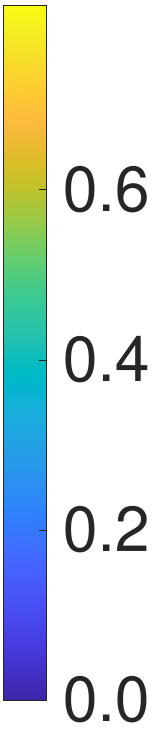}\vspace{-0.2cm}\\
				$M=5$&&$M=10$&&$M=20$&
				\end{tabular}\\
				\begin{tabular}{@{}c@{~~~~}c@{}c@{}}	
				{\raisebox{1.90cm}{\rotatebox[origin=c]{90}{Parafilm }}}&
				\includegraphics[height=0.21\textwidth]{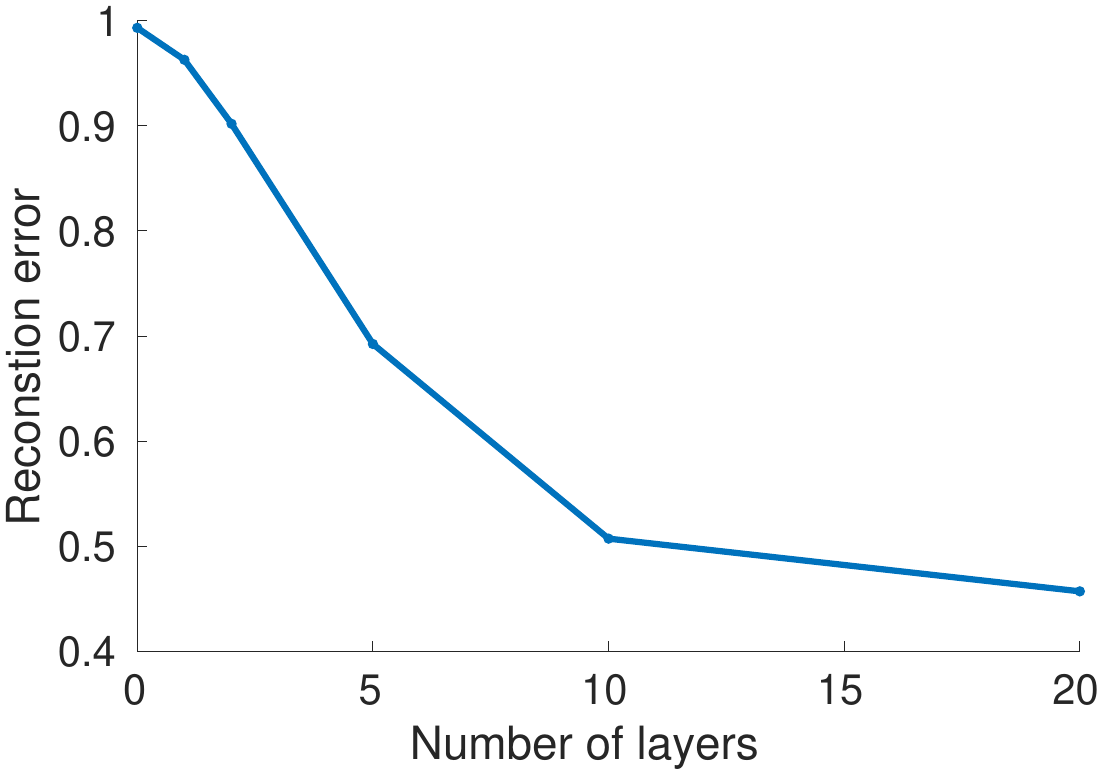}&
				\includegraphics[height=0.21\textwidth]{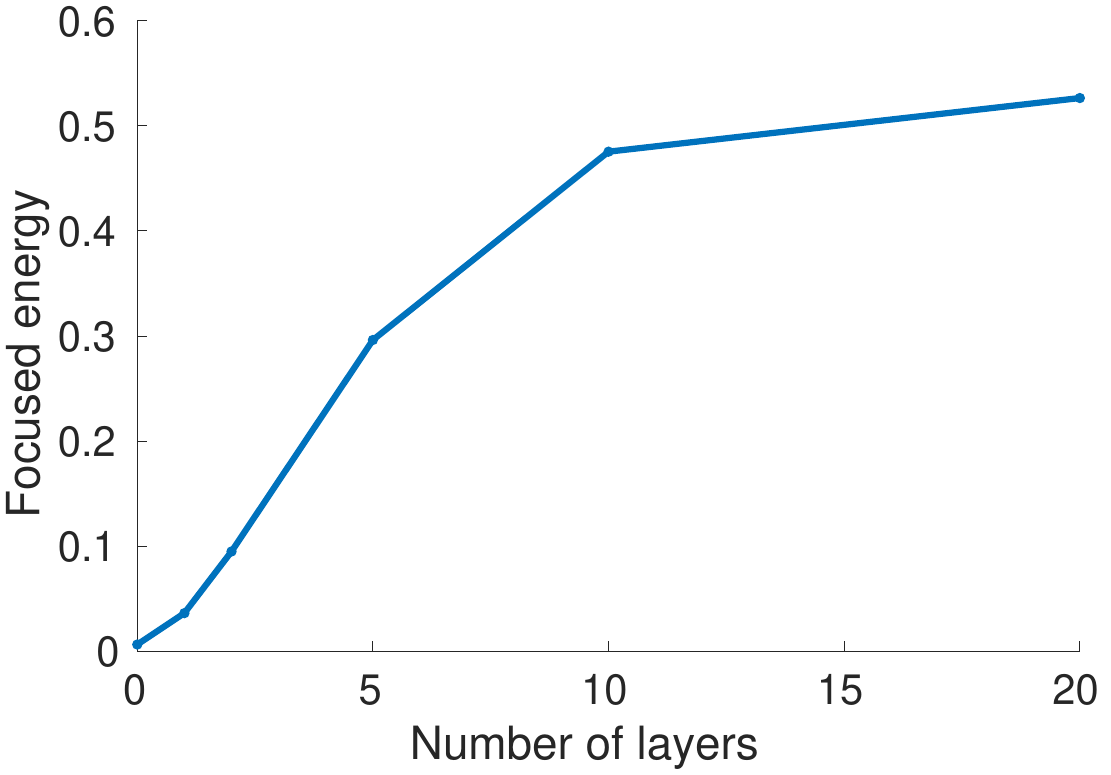}
			\end{tabular}&
			\begin{tabular}{@{}c@{}c@{}c@{}@{}c@{}c@{}c@{}}	
				\includegraphics[height=0.08\textwidth]{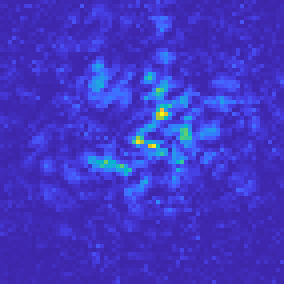}&
				\includegraphics[height=0.08\textwidth]{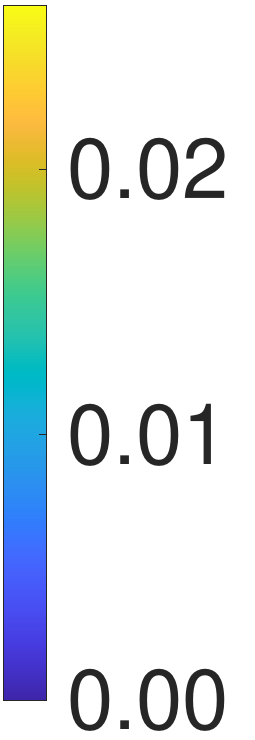}&
				\includegraphics[height=0.08\textwidth]{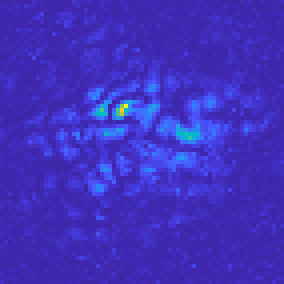}&
				\includegraphics[height=0.08\textwidth]{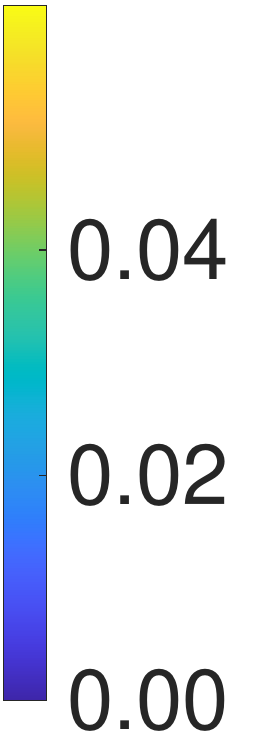}&
				\includegraphics[height=0.08\textwidth]{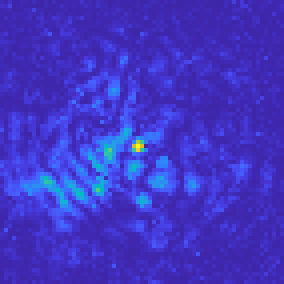}&
				\includegraphics[height=0.08\textwidth]{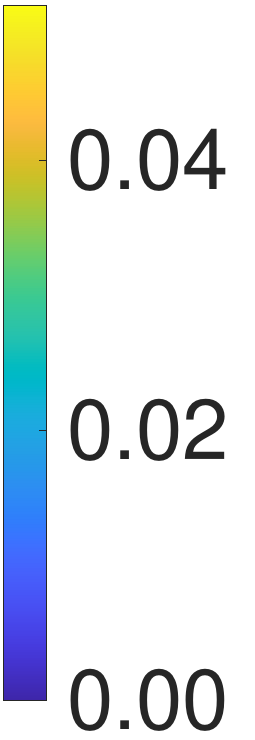}\vspace{-0.2cm}\\
				$M=0$&&$M=1$&&$M=2$&\\
				\includegraphics[height=0.08\textwidth]{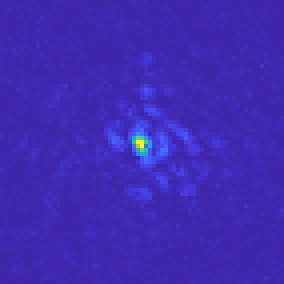}&
				\includegraphics[height=0.08\textwidth]{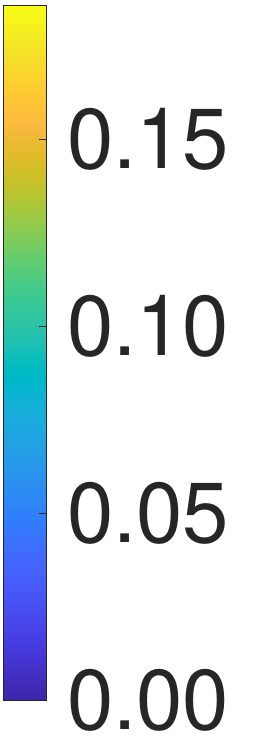}&
				\includegraphics[height=0.08\textwidth]{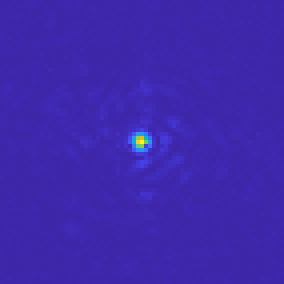}&
				\includegraphics[height=0.08\textwidth]{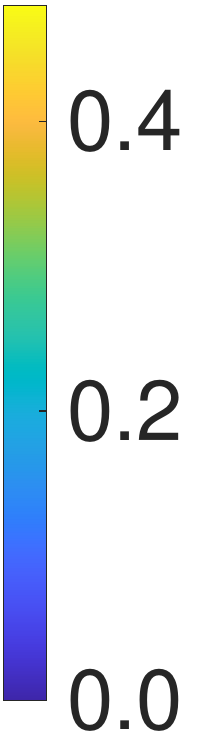}&
				\includegraphics[height=0.08\textwidth]{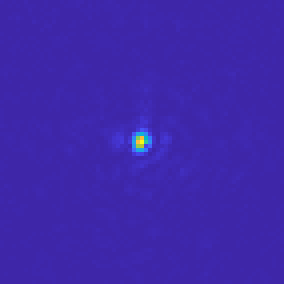}&
				\includegraphics[height=0.08\textwidth]{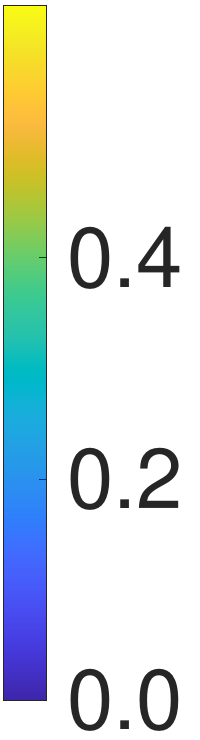}\vspace{-0.2cm}\\
				$M=5$&&$M=10$&&$M=20$&\\	
			\end{tabular}
		\\
			\begin{tabular}{@{}c@{~~~~}c@{}c@{}}	
			{\raisebox{1.90cm}{\rotatebox[origin=c]{90}{Brain}}}&
			\includegraphics[height=0.21\textwidth]{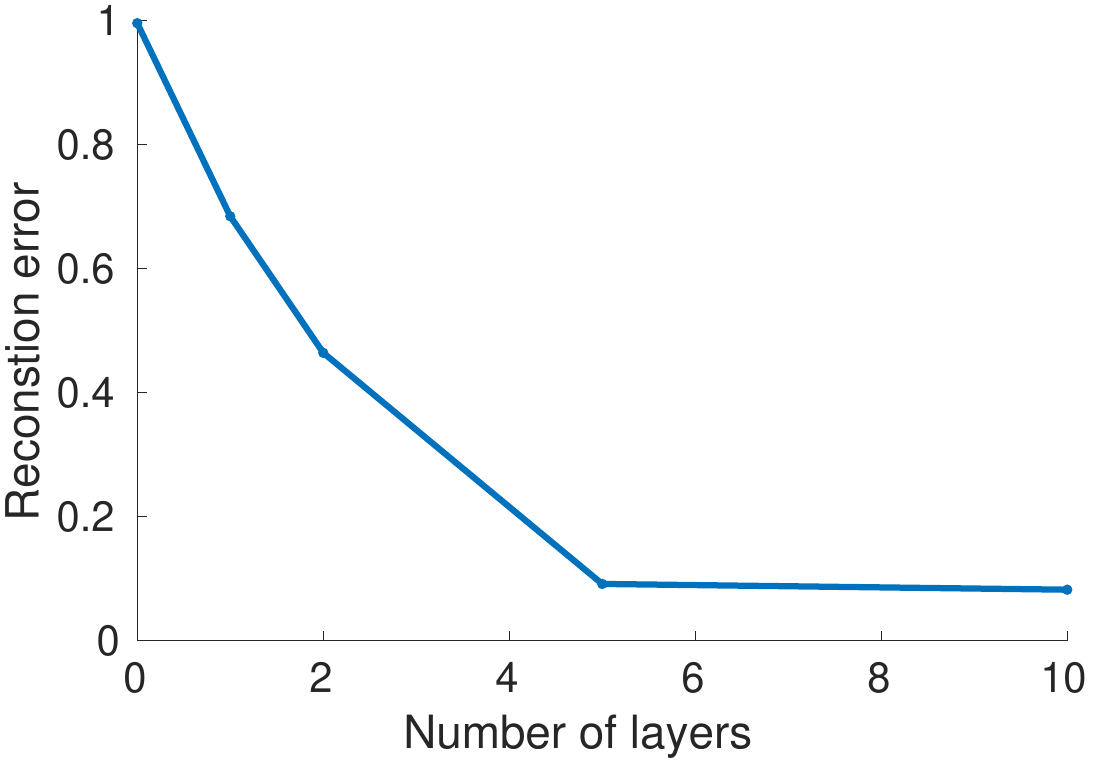}&
			\includegraphics[height=0.21\textwidth]{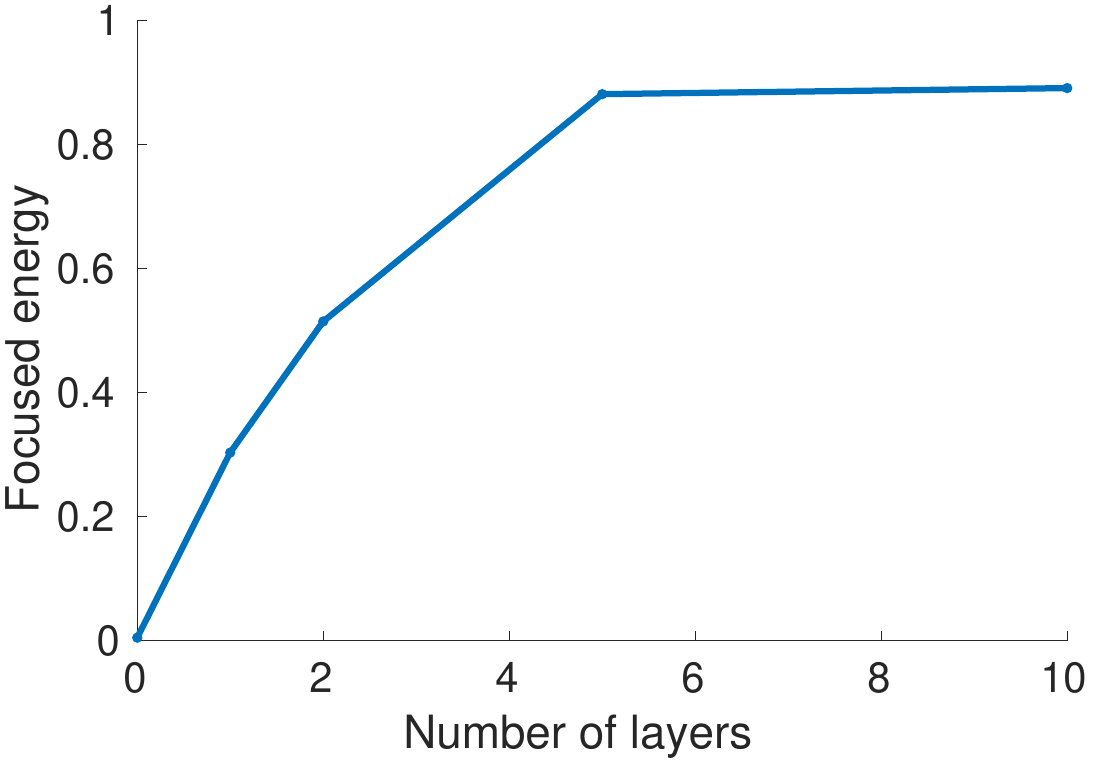}\\
			&	(a) Fitting error&(b) Delivered Eng.
		\end{tabular}&
		\begin{tabular}{@{}c@{}c@{}c@{}@{}c@{}c@{}c@{}}	
			\includegraphics[height=0.08\textwidth]{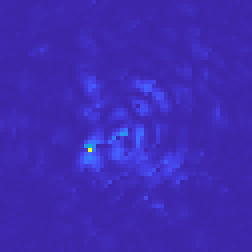}&
			\includegraphics[height=0.08\textwidth]{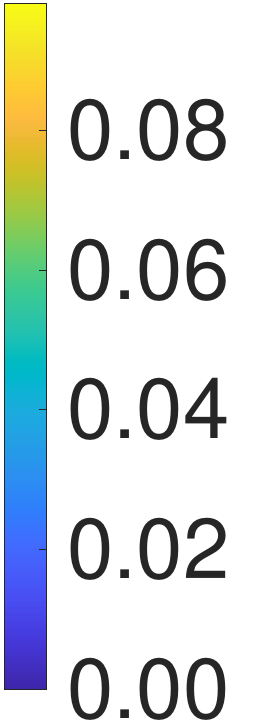}&
			\includegraphics[height=0.08\textwidth]{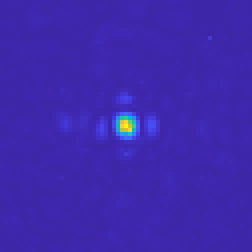}&
			\includegraphics[height=0.08\textwidth]{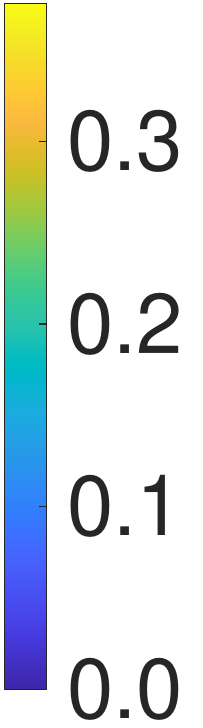}&
			\includegraphics[height=0.08\textwidth]{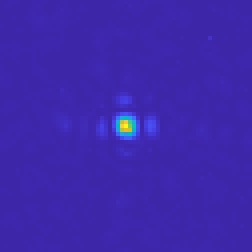}&
			\includegraphics[height=0.08\textwidth]{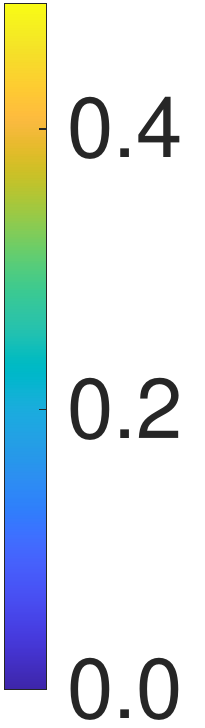}\vspace{-0.2cm}\\
			$M=0$&&$M=1$&&$M=2$&\\
			\includegraphics[height=0.08\textwidth]{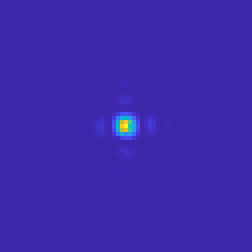}&
			\includegraphics[height=0.08\textwidth]{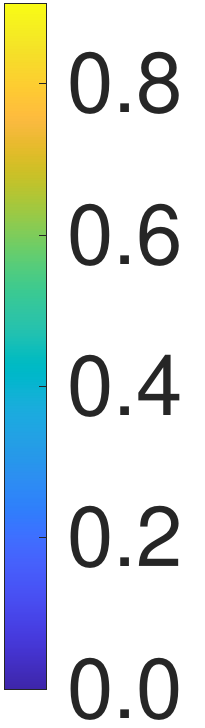}&
			\includegraphics[height=0.08\textwidth]{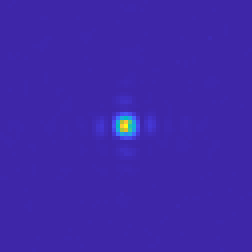}&
			\includegraphics[height=0.08\textwidth]{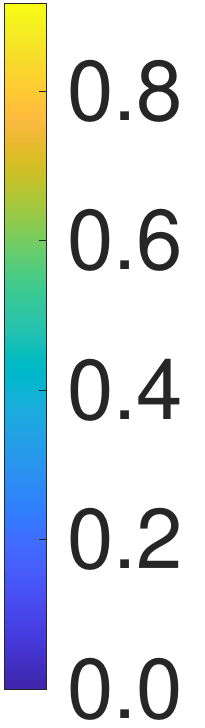}&&
			\\
			$M=5$&&$M=10$&\\	\multicolumn{6}{c}{(c) Focused spots}		
		\end{tabular}
		\end{tabular}
		\caption{\footnotesize{\bf Fitting  transmission matrix:} 
			We fit a few types of transmission matrices with multi-slice model. The top panel tested physically accurate transmission matrices simulated with the Monte-Carlo approach of ~\cite{Bar:2020} at two optical depths.  Lower panels used a transmission matrices measured in the lab through layers of parafilm and mouse brain.    (a)  The reduction in fitting error as a function of the number of layers. (b) The averaged delivered energy (correlation between captured and fitted matrices). (c) 
			An example of a spot behind the tissue using the wavefront estimated by the layered model, with  different number of layers. (for the M.C. matrices we only show focusing at $\od=5$).}\label{fig:fitting-all}
	\end{center}
\end{figure*}

%% file: cvpr_sec_files_cr/fig_fit_chicken_varysup_supp.tex
\begin{figure*}[t!]
	\begin{center}
		\begin{tabular}{@{}c@{~~~~~~~}c@{}}	
		
				\begin{tabular}{@{}c@{}}		
				\includegraphics[height=0.24\textwidth]{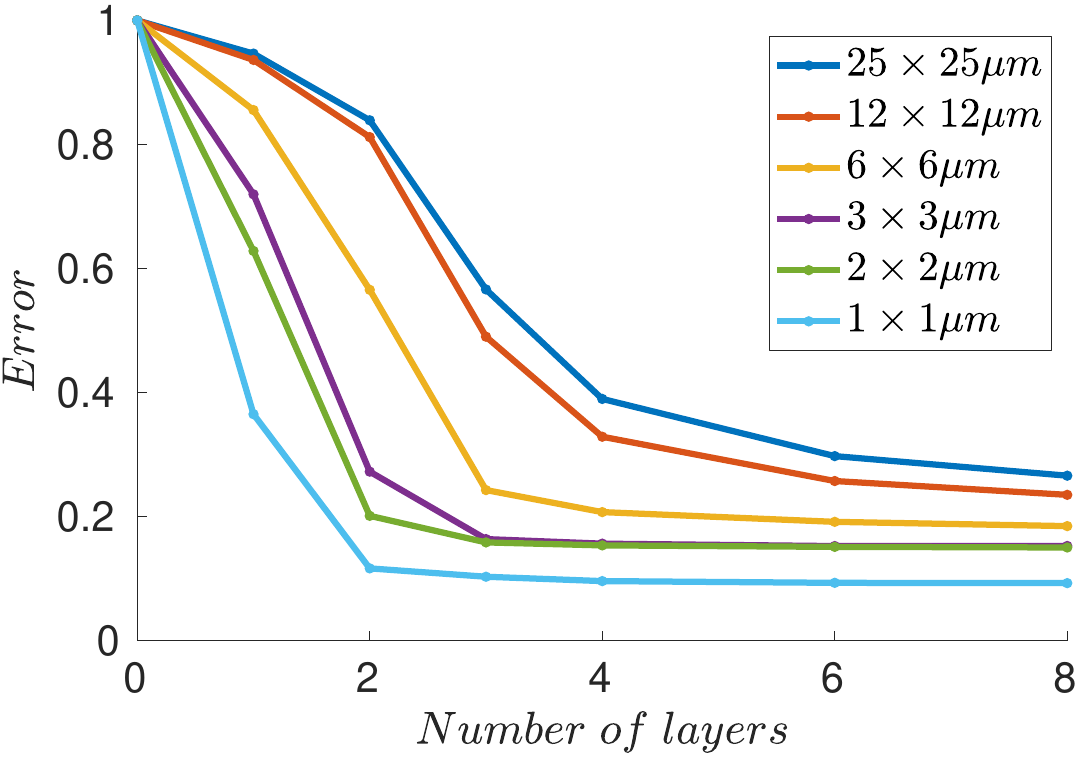}
			\end{tabular}&
			\begin{tabular}{@{}c@{}c@{}c@{}@{}c@{}c@{}c@{}@{}c@{}c@{}c@{}@{}c@{}c@{}c@{}c@{}c@{}}	
				{\raisebox{0.60cm}{\rotatebox[origin=c]{90}{$25\mu m$ }}}&
				\includegraphics[height=0.08\textwidth]{figs/fitting_marina_chicken_varysup/varysup_marina05_30_mats_z_max140_focusing_layerNum0_supind1}&
				\includegraphics[height=0.08\textwidth]{figs/fitting_marina_chicken_varysup/varysup_marina05_30_mats_z_max140_focusing_layerNum0_supind1_colorbar}&		
               \includegraphics[height=0.08\textwidth]{figs/fitting_marina_chicken_varysup/varysup_marina05_30_mats_z_max140_focusing_layerNum1_supind1}&
               \includegraphics[height=0.08\textwidth]{figs/fitting_marina_chicken_varysup/varysup_marina05_30_mats_z_max140_focusing_layerNum1_supind1_colorbar}&		
               \includegraphics[height=0.08\textwidth]{figs/fitting_marina_chicken_varysup/varysup_marina05_30_mats_z_max140_focusing_layerNum2_supind1}&
               \includegraphics[height=0.08\textwidth]{figs/fitting_marina_chicken_varysup/varysup_marina05_30_mats_z_max140_focusing_layerNum2_supind1_colorbar}&		
               \includegraphics[height=0.08\textwidth]{figs/fitting_marina_chicken_varysup/varysup_marina05_30_mats_z_max140_focusing_layerNum3_supind1}&
               \includegraphics[height=0.08\textwidth]{figs/fitting_marina_chicken_varysup/varysup_marina05_30_mats_z_max140_focusing_layerNum3_supind1_colorbar}&				
               \includegraphics[height=0.08\textwidth]{figs/fitting_marina_chicken_varysup/varysup_marina05_30_mats_z_max140_focusing_layerNum6_supind1}&
               \includegraphics[height=0.08\textwidth]{figs/fitting_marina_chicken_varysup/varysup_marina05_30_mats_z_max140_focusing_layerNum6_supind1_colorbar}
               \\
               		{\raisebox{0.60cm}{\rotatebox[origin=c]{90}{$6\mu m$ }}}&
               	\includegraphics[height=0.08\textwidth]{figs/fitting_marina_chicken_varysup/varysup_marina05_30_mats_z_max140_focusing_layerNum0_supind3}&
               \includegraphics[height=0.08\textwidth]{figs/fitting_marina_chicken_varysup/varysup_marina05_30_mats_z_max140_focusing_layerNum0_supind3_colorbar}&		
               \includegraphics[height=0.08\textwidth]{figs/fitting_marina_chicken_varysup/varysup_marina05_30_mats_z_max140_focusing_layerNum1_supind3}&
               \includegraphics[height=0.08\textwidth]{figs/fitting_marina_chicken_varysup/varysup_marina05_30_mats_z_max140_focusing_layerNum1_supind3_colorbar}&		
               \includegraphics[height=0.08\textwidth]{figs/fitting_marina_chicken_varysup/varysup_marina05_30_mats_z_max140_focusing_layerNum2_supind3}&
               \includegraphics[height=0.08\textwidth]{figs/fitting_marina_chicken_varysup/varysup_marina05_30_mats_z_max140_focusing_layerNum2_supind3_colorbar}&		
               \includegraphics[height=0.08\textwidth]{figs/fitting_marina_chicken_varysup/varysup_marina05_30_mats_z_max140_focusing_layerNum3_supind3}&
               \includegraphics[height=0.08\textwidth]{figs/fitting_marina_chicken_varysup/varysup_marina05_30_mats_z_max140_focusing_layerNum3_supind3_colorbar}&				
               \includegraphics[height=0.08\textwidth]{figs/fitting_marina_chicken_varysup/varysup_marina05_30_mats_z_max140_focusing_layerNum6_supind3}&
               \includegraphics[height=0.08\textwidth]{figs/fitting_marina_chicken_varysup/varysup_marina05_30_mats_z_max140_focusing_layerNum6_supind3_colorbar}
               \\
               	{\raisebox{0.60cm}{\rotatebox[origin=c]{90}{$1\mu m$ }}}&
               	\includegraphics[height=0.08\textwidth]{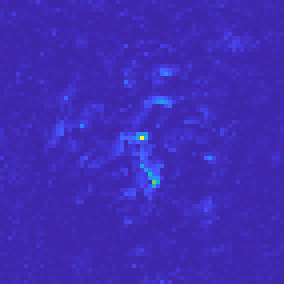}&
               \includegraphics[height=0.08\textwidth]{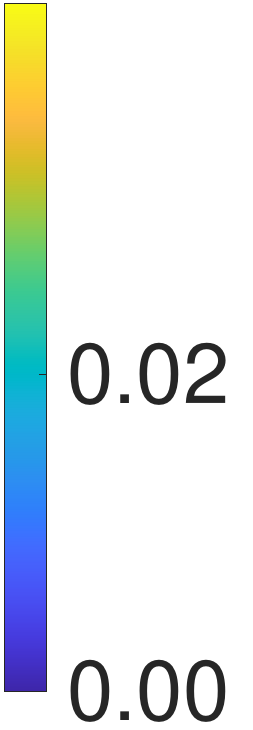}&		
               \includegraphics[height=0.08\textwidth]{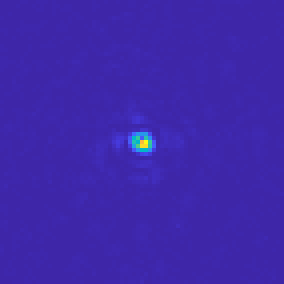}&
               \includegraphics[height=0.08\textwidth]{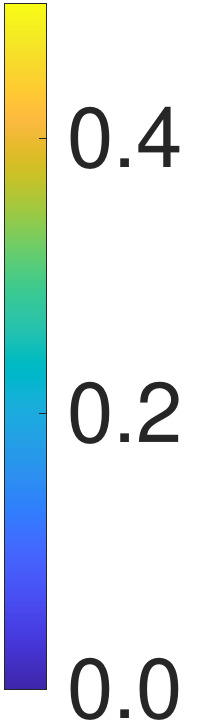}&		
               \includegraphics[height=0.08\textwidth]{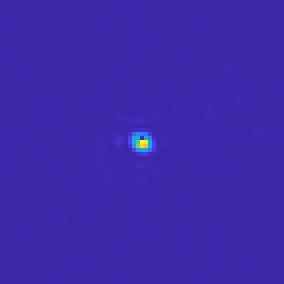}&
               \includegraphics[height=0.08\textwidth]{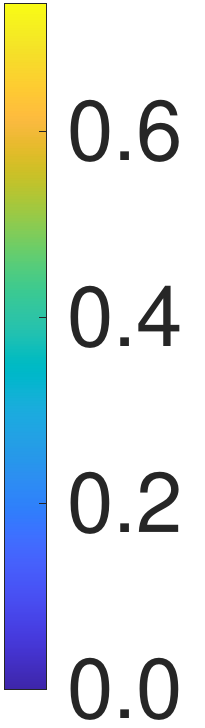}&		
               \includegraphics[height=0.08\textwidth]{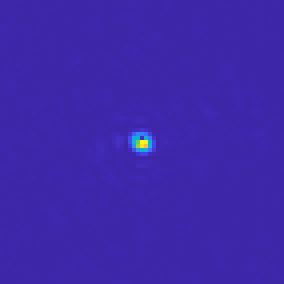}&
               \includegraphics[height=0.08\textwidth]{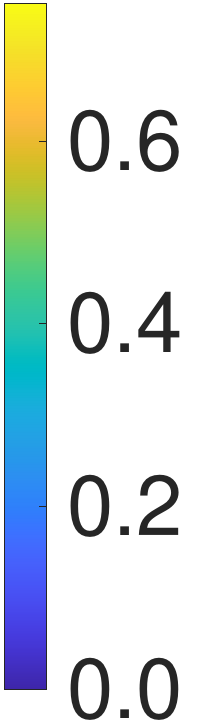}&			
               \includegraphics[height=0.08\textwidth]{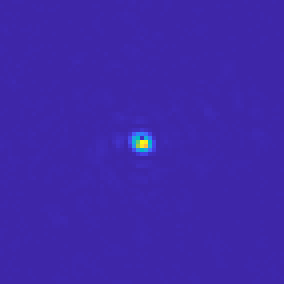}&
               \includegraphics[height=0.08\textwidth]{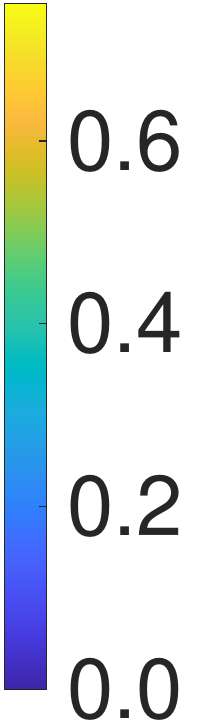}
               \\
               &$M=0$&&$M=1$&&$M=2$&&$M=3$&&$M=6$
			\end{tabular}
		\end{tabular}\vspace{-0.2cm}

\caption{\footnotesize {\bf Layer support for a captured transmission matrix:} We consider a transmission matrix captured in the lab through a chicken breast layer of thickness $170\mu m$. Left:  plotting the fitting error as a function of the number of layers.   Smaller supports can be fitted with fewer layers.    Right: visualization of a spot focusing behind the tissue, computed using the wavefronts of the fitted model.  We compare models fitted to different supports. In the top row we need a larger number of layers, but the spot can be scanned over a $25\times 25\mu m$ window using the same layers. In the lower rows we achieve a focused spot using a smaller number of layers, but these layers can scan the focused spot over smaller windows of sizes $6\times 6\mu m$ and $1\times 1\mu m$.	}\vspace{-0.3cm}
	\label{fig:fitting-chicken-varysup}
	\end{center}
\end{figure*}

%% file: cvpr_sec_files_cr/volumesampling.tex
\section{Volume   synthesis}
To sample the 3D volumes used to simulate  multi-layer transmission matrices we sampled spheres at random positions. We sampled the spheres so that the resulting volume would have a target optical depth $\od$. To this end we had to select two parameters, the sphere density, and the variation in refractive index induced by the sphere.

The sphere radii were uniformly sampled in the range $0.5 - 1.5 \mu m$ (so diameters in the range $1-3\mu m$) with illumination wavelength $\lambda=0.5\mu m$.
We note that a sphere with radii $\varsigma$ has a 2D cross section area of $\pi \varsigma^2$. Therefore given a target optical depth $\od$ the average number of spheres in a volume of size $W\times W\times d$ should be
\BE
\od \frac{ W^2 }{\int p(\varsigma)\pi \varsigma^2 d\varsigma}
\EE
where $p(\varsigma)$ is the probability of sampling a sphere with radii $\varsigma$.

Each sampled sphere is assigned a uniform refractive index $n$, which differs from the  refractive index of the leading medium  by a random value in the range \BE\label{eq:alpha-rng}[-\alpha, \alpha].\EE We select $\alpha$ so that the resulting volume meets the target optical depth as explained next.
To that end we note that the phase masks $\rho_m$ of the multi-slice model in Eq.~2  of the main paper  
are equivalent to the integral of refractive indices variation in a slice of thickness $\eps$ around it. As the refractive index variation has a low magnitude we can approximate the phase mask as

\BE 
\rho_m(x,y)=e^{\frac{2\pi i}{\lambda}n(x,y)}\approx 1 -\left(\frac{2\pi n(x,y)}{\lambda}\right)^2+i \frac{2\pi n(x,y)}{\lambda}
\EE
We approximate $\rho_m$ as
\BE
\rho_m(x,y)=\mu+\delta \rho_m(x,y)
\EE
such that $\mu$ is the mean of $\rho$ (which is a real positive scalar) and $\delta \rho_m(x,y)$ is a zero mean residual.
We note that the mean $\mu$ is effected by the range of refractive indices $\alpha$ in \equref{eq:alpha-rng}. 
On the other hand we note that light propagating through $M$ aberration layers maintains a ballistic component  attenuated by $\mu^M$. To meet a target optical depth we want $\mu^M=e^{-\od}$. We therefore numerically scan multiple values of $\alpha$ and select the one providing the desired optical depth.